\documentclass[twocolumn]{aastex631}

\begin{document}

\title{High-contrast JWST-MIRI spectroscopy of planet-forming disks for the JDISC Survey}

\correspondingauthor{Klaus M. Pontoppidan}
\email{klaus.m.pontoppidan@jpl.nasa.gov}
\author[0000-0001-7552-1562]{Klaus M. Pontoppidan}
\affiliation{Jet Propulsion Laboratory, California Institute of Technology, 4800 Oak Grove Drive, Pasadena, CA 91109, USA}
\affiliation{Division of Geological and Planetary Sciences, California Institute of Technology, MC 150-21, 1200 E California Boulevard, Pasadena, CA 91125, USA}

\author[0000-0003-3682-6632]{Colette Salyk}
\affil{Department of Physics and Astronomy, Vassar College, 124 Raymond Avenue, Poughkeepsie, NY 12604, USA}

\author[0000-0003-4335-0900]{Andrea Banzatti}
\affil{Department of Physics, Texas State University, 749 N Comanche Street, San Marcos, TX 78666, USA}

\author[0000-0002-0661-7517]{Ke Zhang}
\affil{Department of Astronomy, University of Wisconsin-Madison, Madison, WI 53706, USA}

\author[0000-0001-7962-1683]{Ilaria Pascucci}
\affil{Department of Planetary Sciences, University of Arizona, 1629 East University Boulevard, Tucson, AZ 85721, USA}

\author[0000-0001-8798-1347]{Karin I. \"Oberg}
\affiliation{Center for Astrophysics, Harvard \& Smithsonian, 60 Garden St., Cambridge, MA 02138, USA}

\author[0000-0002-7607-719X]{Feng Long}
\affiliation{Lunar and Planetary Laboratory, University of Arizona, Tucson, AZ 85721, USA}

\author[0000-0001-7152-9794]{Carlos Mu\~{n}oz-Romero}
\affiliation{Center for Astrophysics, Harvard \& Smithsonian, 60 Garden St., Cambridge, MA 02138, USA}

\author[0000-0002-6695-3977]{John Carr}
\affiliation{Department of Astronomy, University of Maryland, College Park, MD 20742, USA}

\author[0000-0002-5758-150X]{Joan Najita}
\affiliation{NSF’s NOIRLab, 950 N. Cherry Avenue, Tucson, AZ 85719, USA}

\author[0000-0003-0787-1610]{Geoffrey A. Blake}
\affiliation{Division of Geological and Planetary Sciences, California Institute of Technology, MC 150-21, 1200 E California Boulevard, Pasadena, CA 91125, USA}

\author[0000-0003-2631-5265]{Nicole Arulanantham}
\affiliation{Space Telescope Science Institute, 3700 San Martin Drive, Baltimore, MD 21218, USA}

\author[0000-0003-2253-2270]{Sean Andrews}
\affiliation{Center for Astrophysics, Harvard \& Smithsonian, 60 Garden St., Cambridge, MA 02138, USA}

\author[0000-0002-4276-3730]{Nicholas P. Ballering}
\affiliation{Department of Astronomy, University of Virginia, Charlottesville, VA 22904, USA}

\author[0000-0003-4179-6394]{Edwin Bergin}
\affiliation{Department of Astronomy, University of Michigan, Ann Arbor, MI 48109, USA}

\author[0000-0002-0150-0125]{Jenny Calahan}
\affiliation{Center for Astrophysics, Harvard \& Smithsonian, 60 Garden St., Cambridge, MA 02138, USA}

\author{Douglas Cobb}
\affil{Department of Physics and Astronomy, Vassar College, 124 Raymond Avenue, Poughkeepsie, NY 12604, USA}

\author[0000-0002-5296-6232]{Maria Jose Colmenares}
\affiliation{Department of Astronomy, University of Michigan, Ann Arbor, MI 48109, USA}

\author[0000-0002-4555-5144]{Annie Dickson-Vandervelde}
\affil{Department of Physics and Astronomy, Vassar College, 124 Raymond Avenue, Poughkeepsie, NY 12604, USA}

\author{Anna Dignan}
\affil{Department of Physics and Astronomy, Vassar College, 124 Raymond Avenue, Poughkeepsie, NY 12604, USA}

\author[0000-0003-1665-5709]{Joel Green}
\affiliation{Space Telescope Science Institute, 3700 San Martin Drive, Baltimore, MD 21218, USA}

\author[0009-0008-3735-4124]{Phoebe Heretz}
\affil{Department of Physics and Astronomy, Vassar College, 124 Raymond Avenue, Poughkeepsie, NY 12604, USA}

\author[0000-0002-7154-6065]{Greg Herczeg}
\affiliation{Kavli Institute for Astronomy and Astrophysics, Peking University, Yi He Yuan Lu 5, Haidian Qu, 100871 Beijing, China}

\author[0000-0002-5067-1641]{Anusha Kalyaan}
\affiliation{Department of Physics, Texas State University, 749 N Comanche Street, San Marcos, TX 78666, USA}

\author[0000-0002-3291-6887]{Sebastiaan Krijt}
\affiliation{School of Physics and Astronomy, University of Exeter, Stocker Road, Exeter EX4 4QL, UK}

\author[0000-0001-9500-9267]{Tyler Pauly}
\affiliation{Space Telescope Science Institute, 3700 San Martin Drive, Baltimore, MD 21218, USA}

\author[0000-0001-8764-1780]{Paola Pinilla}
\affiliation{Mullard Space Science Laboratory, University College London, Holmbury St Mary, Dorking, Surrey RH5 6NT, UK}

\author[0000-0002-8623-9703]{Leon Trapman}
\affil{Department of Astronomy, University of Wisconsin-Madison, Madison, WI 53706, USA}

\author[0000-0001-8184-5547]{Chengyan Xie}
\affiliation{Lunar and Planetary Laboratory, University of Arizona, Tucson, AZ 85721, USA}

\begin{abstract}
The JWST Disk Infrared Spectral Chemistry Survey (JDISCS) aims to understand the evolution of the chemistry of inner protoplanetary disks using the Mid-InfraRed Instrument (MIRI) on the James Webb Space Telescope (JWST). With a growing sample of $>30$ disks, the survey implements a custom method to calibrate the MIRI Medium Resolution Spectrometer (MRS) to contrasts of better than 1:300 across its 4.9-28\,$\mu$m spectral range. This is achieved using observations of Themis-family asteroids as precise empirical reference sources. High spectral contrast enables precise retrievals of physical parameters, searches for rare molecular species and isotopologues, and constraints on the inventories of carbon- and nitrogen-bearing species. JDISCS also offers significant improvements to the MRS wavelength and resolving power calibration. We describe the JDISCS calibrated data and demonstrate its quality using observations of the disk around the solar-mass young star FZ Tau. The FZ Tau MIRI spectrum is dominated by strong emission from warm water vapor. We show that the water and CO line emission originates from the disk surface and traces a range of gas temperatures of $\sim$500--1500\,K. We retrieve parameters for the observed CO and H$_2$O lines, and show that they are consistent with a radial distribution represented by two temperature components. A high water abundance of $n({\rm H_2O})\sim 10^{-4}$ fills the disk surface at least out to the 350\,K isotherm at 1.5\,au. We search the FZ Tau environs for extended emission detecting a large (radius of $\sim$300\,au) ring of emission from H$_2$ gas surrounding FZ Tau, and discuss its origin. 
\end{abstract}


\section{Introduction} 
Giant planets, planetesimals, the embryos of terrestrial planets, as well as comets and Kuiper belt objects, form in a rich chemical environment where primordial water and other volatiles mix with new complex chemistry to create a vast diversity of planetary systems \citep{Oberg23}. In analogy with our own solar system, rocky planets probably have a wide range of compositions, depending on their formation history. Some may be rich in ices and volatiles, whereas others are rocky with tenuous atmospheres. While the formation and chemical evolution of planets is complex, it is widely thought that the initial chemical conditions of oxygen, carbon, and nitrogen of the planet-forming region at 1-10 AU \citep{Mordasini09} play a vital role in determining the final makeup of planets \citep{Cleeves14, Bergin15}. Indeed, many chemical signatures seen in ancient Solar System material likely have their origin in the gas-rich protoplanetary disk phase \citep{Busemann06,Mumma11}. 

\subsection{Tracing the chemistry of planet-forming gas with infrared spectroscopy}
While millimeter-wave observatories, such as ALMA, are powerful facilities for tracing cold molecules at large disk radii (10-100 AU), many of the most abundant (bulk) molecular species in planet-forming regions, such as water, CO$_2$, and CH$_4$, can only be efficiently characterized at infrared wavelengths from space. Extensive spectroscopic observations obtained with the InfraRed Spectrometer (IRS) on Spitzer discovered that many protoplanetary disks surrounding low-mass young stars (up to about 1.5\,$M_{\odot}$) have mid-infrared spectra covered in bright emission lines from warm molecular gas \citep{Carr08,Salyk08,Pontoppidan10,Carr11,Pascucci13}. It has further been shown that the gas emitting at infrared wavelengths originates at radii of 0.1-10 AU, commensurate with the primary planet-forming region \citep{Salyk11}. 

Because of the high dust optical depth of typical inner disks, the infrared emission generally traces the surface, representing 0.1--1\% of the total vertical column density at 1\,au \citep{Woitke18,bosman22}. These regions are expected to have highly active chemistry due to the high temperatures and exposure to UV and X-ray radiation from the central star and accretion shock \citep{Semenov11,Calahan22}. Consequently, molecular abundances may be strongly altered, by orders of magnitude, from their primordial composition \citep{Pontoppidan14}. Indeed, a combination of gas-phase kinetic reactions of self-shielding is expected to lead to abundant water vapor in the inner disk at temperatures above $\sim$300\,K \citep{Bethell09}. Concurrently, such static models predict that cooler gas (150-300\,K) is deficient in water, resulting in observable column densities below $\sim 10^{15}\,\rm cm^{-2}$. 

However, this simplified static chemistry may be strongly modified by a combination of selective freeze-out and vertical/radial drift of millimeter to meter-sized pebbles \citep{Meijerink09, Kama16, Booth19, Banzatti20, Price22}. Since water is a potentially dominant solid mass reservoir, such transport may have profound effects on the formation of planetesimals and planetary cores close to the snowline.

A massive flow of icy pebbles across the snowline, followed by vertical turbulent mixing, may enhance the abundance of water in gas at temperatures below $\sim$300\,K by orders of magnitude above the low abundances predicted by static chemical models. Recently, JWST has begun confirming the Spitzer observations of molecular emission throughout the mid-infrared \citep{Banzatti23}, and initial results indeed suggest a wide diversity in relative molecular abundances \citep{Grant23, Tabone23, Gasman23}. 

\subsection{The need for high-contrast mid-infrared spectra}
These fundamental questions motivate a strong need for obtaining high SNR mid-infrared spectra of protoplanetary disks in order to: 1) Precisely measure the water vapor column density via the detection of optically thin lines, 2) detect rare molecular species, including isotopologues, and 3) place limits on potential carriers of the most abundant elements, such as NH$_3$ for nitrogen and CH$_4$ for carbon. Precisely measuring the abundance of the H$_2^{16}$O at a range of temperatures requires the detection of weak lines because the strongest water lines are highly optically thick \citep{Meijerink09,Notsu16}, and trace the emitting area, rather than the total column density. The optically thin lines have small intrinsic strength ($A_{ul}\lesssim 10^{-3}\,\rm s^{-1}$) and may be 10--100 times weaker than the brightest water lines. However, since the mid-infrared Spectral Energy Distributions (SEDs) of protoplanetary disks are typically characterized by bright dust emission, searches for rare molecular species and optically thin lines tracing the total water column density often require very high signal-to-noise ratios and correspondingly high spectral contrast. Further, detecting rare isotopologues of water and other molecules would allow us to measure the H$_2^{16}$O/H$_2^{18}$O ratio as a proxy for photochemical processing that is believed to have occurred in the early solar system  \citep[e.g.,][]{Clayton73,Lyons05}. Important nitrogen- and carbon-bearing species are expected to be present at low contrast. HCN and C$_2$H$_2$ are typically detected via their Q-branches near 14\,$\mu$m, but accurate column density measurements require the use of the Q-branch shape, in combination with the much weaker R and P branches. Other rare species are also important tracers of inner disk chemistry and evolution. An active inner disk chemistry will lead to a re-distribution of volatile elements into new molecular carriers. For instance, NH$_3$ -- a major nitrogen carrier in the cold interstellar medium -- is likely destroyed in the disk surface, and the nitrogen driven into N$_2$ and HCN \citep{Walsh15}, which could be part of the explanation for why the Earth is so depleted in nitrogen. We can use the strong $\nu_2$ rovibrational band of warm NH$_3$ around 10\,$\mu$m to detect, or put strong constraints on the inner disk NH$_3$ abundance, down to abundances of $\sim 10^{-8}$, relative to H$_2$ \citep{Pontoppidan19}. 

Together, all these species allow constraints of the bulk elemental abundances (C/O and C/N) of the primary planet-forming regions for direct comparisons with observations of these quantities in exoplanetary atmospheres \citep{Oberg11, Madhusudhan12}. 

\begin{deluxetable}{lcl}[ht!]
\tablecaption{FZ Tau star and disk parameters}
\label{table:fztau_pars}
\tablehead{
\colhead{Parameter} & \colhead{Value} &\colhead{Reference}
} 
\startdata
Distance      & $129.2\pm0.4$\,pc & Gaia DR3 \\
SpT           & M0V & \cite{McClure19} \\
$M_*$         & 0.56\,$M_{\odot}$&\cite{McClure19}\\
$L_*$         & 0.97\,$L_{\odot}$&\cite{McClure19}\\
$T_{\rm eff}$ & 3850\,K& \cite{McClure19}\\
$R_*$         & 2.18\,$R_{\odot}$& \cite{McClure19} \\
$M_{\rm acc}$ &$3.5\times 10^{-7}\,M_{\odot}\,yr^{-1}$&  \cite{McClure19} \\
$M_{\rm disk}$ & $1.2-5.0\times 10^{-3}\,M_{\odot}$ & \cite{Akeson19} \\
               &                                    & \cite{Ribas20}\\
$R_{\rm disk, mm-dust}$ &$11.6^{+3.0}_{-1.2}$\,au& \cite{Ribas20} \\
$i$            & $\sim$ 25\degr & F. Long, Priv. Comm. 
\enddata
\end{deluxetable}
\vspace{-1.5cm}

\subsection{The JDISC Survey}
The JWST Disk Infrared Spectral Chemistry Survey (hereinafter JDISCS) is a collaboration involving several JWST programs to observe a well-characterized sample of protoplanetary disks with the Medium Resolution Spectrometer \citep[MRS,][]{Wells15} on the JWST Mid-Infrared Spectrometer \citep[MIRI,][]{Rieke15} at high spectral contrast, reduced using a consistent process. Currently, JDISCS includes PID1549 (PI:K. Pontoppidan), PID1584 (CoPIs: C. Salyk, K. Pontoppidan), PID1640 (PI: A. Banzatti), and PID3034 (PI: K. Zhang). In this paper, we present the data and processing strategies developed for JDISCS and quantify the data quality. A key feature is the use of asteroids to remove the well-known MIRI detector fringes at high signal-to-noise across the full MIRI-MRS wavelength range of 4.9-28\,$\mu$m. This approach is described in Sections \ref{section:strategy} and \ref{section:data_reduction}. In section \ref{section:contrast}, we demonstrate that we achieve the goal of $\ge$1:300 spectral contrast. 

As a demonstration of common use cases for the JDISCS data, we present the MIRI MRS spectrum of the disk around FZ Tau, a low-mass young star in the nearby Taurus star-forming region. Table \ref{table:fztau_pars} summarizes the physical properties of FZ Tau. FZ Tau has a very compact millimeter-dust disk with a radius of $\sim 11.6^{+3.0}_{-1.2}\,$au, yet with a relatively high mass ($>1.2\times 10^{-3}\,M_{\odot}$, Table \ref{table:fztau_pars}), and high accretion rate \citep{McClure19}. Compact millimeter-dust disks like FZ Tau are predicted to have high column densities of warm water vapor, provided that their compactness is due to efficient inward drift of icy pebbles \citep{Banzatti20,Banzatti23b}. Indeed, FZ Tau is known to exhibit abundant warm water \citep{Pontoppidan10}, and it is one of the few disks around solar-mass stars with sensitive Herschel-PACS spectroscopy of cool water, and with estimates of the water abundances near, and outside the surface water snowline at 3.3$\pm$0.2\,au \citep{Blevins16}. FZ Tau has a high accretion rate, which indicates efficient delivery of material, including water, to its inner disk. 

We organize this paper as follows. In Section \ref{section:strategy} we describe how the JDISCS observations are carried out. Section \ref{section:data_reduction} describes the JDISCS custom data reduction process and quantify its performance. In section \ref{section:analysis}, we provide examples of analyses undertaken by the JDISCS team: We present the JDISCS dust continuum estimator applied to FZ Tau (Section \ref{section:dust}). In Section \ref{section:lines}, we retrieve the properties of molecular gas in the inner planet-forming regions of disks (distribution, abundance, and temperatures) for FZ Tau. In Section \ref{section:extended}, we present continuum-subtracted line images showing extended emission from rotational H$_2$ lines. Finally in Section \ref{section:discussion}, we interpret the retrieved FZ Tau parameters in the context of classical protoplanetary disk models.  

\section{Observations} 

\subsection{High-contrast observing strategy}
\label{section:strategy}
MIRI-MRS spectra suffer from strong, high-frequency fringes due to interference within the detectors with amplitudes of $\sim$10-30\%. This leads to strong confusion with the densely packed molecular spectra from planet-forming disks. While modeling of the fringe response function may achieve contrasts of 1-2\% \citep{Argyriou20}, it is well-known that theoretical fringe fitting methods, such as those used by \cite{Lahuis03} and employed by the MIRI pipeline, are prone to removing power from intrinsic lines by acting as a low-pass filter \citep{Pontoppidan10}. This is particularly problematic if the line contrast is low, or if the line spacing is regular or semi-periodic, as is the case for many molecular bands. Thus, correcting spectra removing them down to a level of 0.3\% requires an independent, direct measurement of the Relative Spectral Response Function (RSRF). The use of empirical calibrators for efficient and stable fringe correction was developed by multiple independent groups for use with observations with the Spitzer Infrared Spectrometer \citep[Spitzer-IRS, ][]{Carr08,Pontoppidan10,Lebouteiller15}, which was affected by similar fringing. Early efforts with MIRI-MRS suggest that similar calibration with standard stars provide promising results at short wavelengths where stars are the brightest \citep{Gasman23_fringe}. 

\begin{figure}[ht!]
\centering
\includegraphics[width=8.5cm]{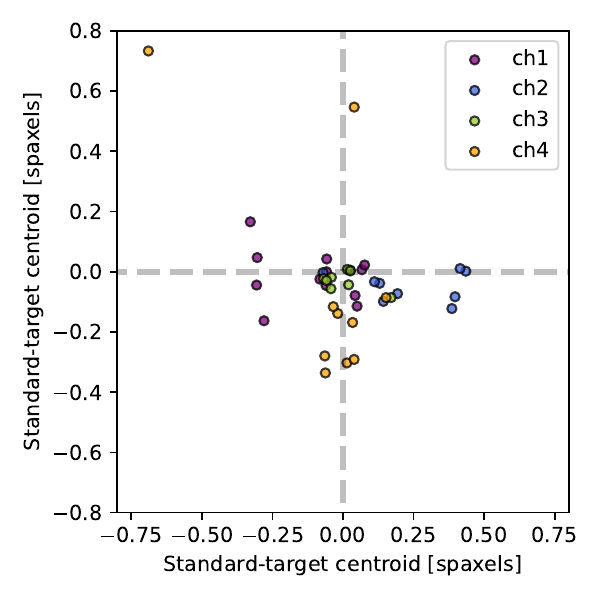}
\caption{Accuracy of the MIRI target acquisition measured as the difference in retrieved centroid for each dither between the asteroid calibrator and the FZ Tau observation. The largest differences, in units of spaxels, is seen for the shortest wavelength channels where the spaxels are the smallest. \label{fig:ta_performance}}
\end{figure}

The observing strategy for the JDISC Survey was designed to optimize the line contrast for high signal-to-noise point sources by obtaining observations of high signal-to-noise calibrators \citep{Pontoppidan21}. The calibrators are required to be bright in the important MRS channel 4, which covers the coolest water lines accessible to JWST, as well as the most unblended H$_2^{18}$O and HDO lines. Given the excellent JWST pointing stability \citep{Rigby23}, even single standard calibrator observations is sufficient for excellent fringe removal of MRS observations. Empirical Relative Spectral Response Functions (RSRFs) are determined independently for each MRS dither position by observing the standard calibrators centered on the relevant spaxels in the MIRI MRS field. The correction is then optimized by placing the science target on the same spaxels, using the same dither pattern. This is a key difference from the JWST calibration pipeline, which does not enforce such ``spaxel matching''. In Figure \ref{fig:ta_performance}, we show that the pointing difference between calibrator and science target are accurate to a fraction of a spaxel, as expected.

\subsection{Asteroid calibrators} 
\label{section:calibrators}
The Cycle 1 MIRI flux calibration plan uses stars for spectrophotometric calibration. These work well at short wavelengths \citep{Gasman23_fringe}, but have much lower SNR in the key Channels 3 and 4 of MIRI-MRS \citep[$\lesssim10$ in MRS Channel 4, ][]{Gordon22}. Consequently, the Cycle 1 JWST calibration program is insufficient to reach the high signal-to-noise needed to fully characterize the molecular spectra from protoplanetary disks. Instead, JDISCS included observations of asteroids as an integral part of PID1549 to provide high SNR fringe calibration across the full MRS range. 

Asteroids are near ideal {\it relative} empirical calibrators for MIRI MRS. They are very bright at 10--30\,$\mu$m, and they generally have spectra that can be approximated by smooth greybodies with shallow broad-band features above a few \% and have no gas-phase lines \citep{Harris98}. Even relatively large asteroids (up to $\sim$ 200\,km diameter) remain point sources when observed with the MIRI MRS beam (corresponding to $\lesssim 0\farcs14$ at a minimum distance to the main asteroid belt of $\sim$2\,au). These are desirable qualities for efficient and model-independent removal of fringes and other spectral response function features in high signal-to-noise MIRI-MRS spectra. The main drawback of asteroids as photometric calibrators is that they are variable by both distance and rotation. Asteroids are therefore used to produce high-quality {\it relative} spectrophotometric calibration of both high- and low-frequency spectral response structure, but not for {\it absolute} photometry. In practice, the latter still requires the additional observation of a well-modeled standard star, such as those provided by the JWST calibration program. 

For measuring the MRS RSRF, we use two Themis family asteroids: 515 Athalia and 526 Jena (see Table \ref{table:asteroid_pars} for their properties). The Themis family consists of primitive C-type asteroids from the outer main belt \citep{Hiriyama18,ziffer11}. A sufficient number of Themis asteroids are available with sizes such that they present a point source to MIRI-MRS with 10-20\,$\mu$m flux densities of 0.2-1.0\,Jy. They are characterized by carbon and volatile-rich surfaces with low silicate content, leading to nearly featureless mid-infrared spectra, with silicate features less than a few \%, as demonstrated by Spitzer spectroscopy \citep{Licandro12}. 526 Jena and 515 Athalia are both spatially unresolved with MIRI:  With diameters of at 43 and 52\,km, respectively, they subtend solid angles of 25--30\,mas at a distance from L2 of $\sim$ 3\,AU, compared to the MIRI MRS resolution of 0.4-1\arcsec.

\begin{figure*}[ht!]
\centering
\includegraphics[width=17cm]{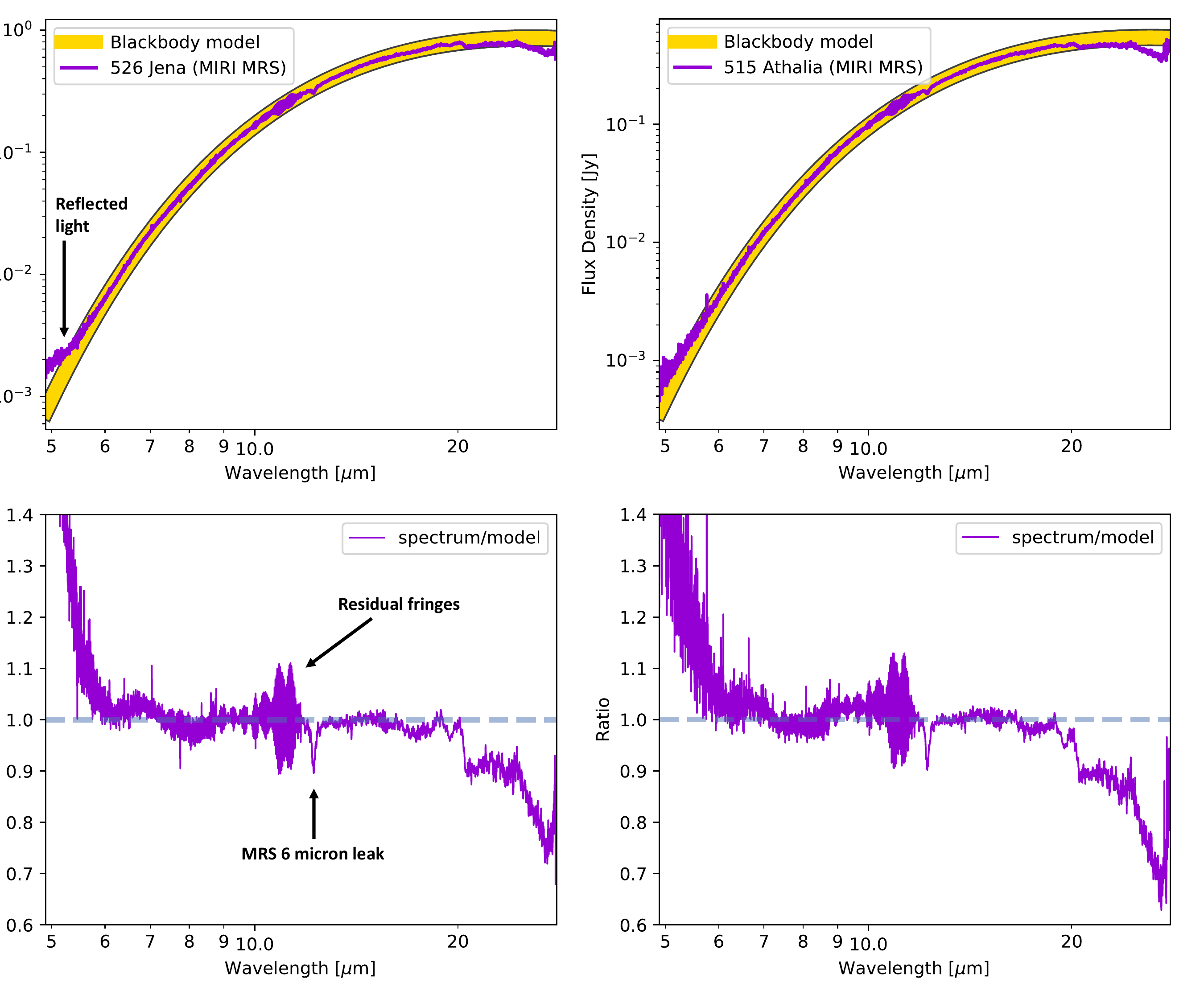}
\caption{Top panels: MIRI MRS spectrum of the Themis family calibrators 526 Jena (left) and 515 Athalia (right), extracted from the stage 2 products using the JWST pipeline spectro-photometric calibration. The spectra are well-fit by blackbody curves using wavelengths above 7\,$\mu$m. Below 7\,$\mu$m, the asteroid spectra become increasingly dominated by reflected light, as opposed to pure thermal emission. However, in this range (Channel 1), we use the standard star HD163466 for the RSRF. Bottom panels: The ratio between spectrum and model. The relative departures between the asteroid and pipeline calibration at long wavelengths are likely due to degradation in the Channel 4 throughputs since commissioning. \label{fig:jena}}
\end{figure*}

We show the asteroid spectra and models used for the calibration in Figure \ref{fig:jena}. It is seen that the asteroid spectrum is highly complementary to observations of stellar calibrators that are bright at short wavelengths. For the JDISCS RSRFs, we supplement the asteroid calibration with MRS observations of the standard calibrator star HD163466 (PID1539) in Channel 1, where the asteroid is faint. 

\subsection{JDISC Survey observations}
\label{section:observations}
The JDISC Survey uses MIRI-MRS to obtain deep spectra of a large sample of protoplanetary disks around $\sim$ 0.1-2 $M_{\odot}$ stars. We observe the JDISCS targets using all three sub-bands (SHORT/A, MEDIUM/B, LONG/C) to cover the full wavelength range between 4.9-28.6\,$\mu$m. The exposure times have a general goal of reaching signal-to-noise ratios (SNR) of at least 300 over the full MRS wavelength range, but in particular at wavelengths longer than 20\,$\mu$m. Exposures are designed to have integration ramps that are as long as possible without saturating the detector to minimize the detector read noise. We consequently used integrations with between 14 and 26 groups, ensuring good slope fits to the ramps, even in the presence of first-frame effects and non-linearity. As presented in this paper, FZ Tau was observed by JWST on February 28, 2023 with exposure times of 14$\times$6 and 21$\times$4 (frames$\times$integrations), yielding 966 and 987 seconds for the short- and long-wavelength detectors, respectively. 

The 4-point ``Negative'' dither, optimized for point sources, samples the RSRF for four specific locations in the MRS field of view. The asteroid calibrators were observed using the same dither pattern to precisely sample the same four point-source fringe patterns. To ensure that the targets and asteroid calibrators are placed on the same spaxel for each dither, all observations were acquired using the MRS target acquisition procedure. This approach uses the neutral density filter to peak up on the brightest pixel of the target source in a region of interest (ROI) in the MIRI imager, before offsetting the source to the MRS field of view. The on-board peak-up algorithm uses the brightest pixel in the target acquisition image to re-center the target in the MRS FOV, which results in centering that is accurate to 0\farcs1, or 0.3-0.5 spaxels, depending on the channel. This precision is verified by our observations, as shown in Figure \ref{fig:ta_performance}. 

\begin{deluxetable}{lc}[ht!]
\tablecaption{JDISCS calibration asteroid parameters (obtained from JPL Horizons or measured from our data).}
\label{table:asteroid_pars}
\tablehead{
\colhead{Parameter} & \colhead{Value} 
}
\startdata
515 Athalia &  \\
\hline
Diameter & $43\pm 0.2$\,km  \\
Albedo & 0.031  \\
Rotation Period & 20.3\,hours  \\
Semimajor axis & 3.09\,au  \\
$F_{\nu}(10\,\mu \rm m)$ & 0.1\,Jy  \\
$F_{\nu}(20\,\mu \rm m)$ & 0.5\,Jy  \\
Surface Temperature & 195$\pm$5\,K\\
Observing date & 2022/11/21  \\
\hline
526 Jena & \\
\hline
Diameter & $52.3\pm 0.5$\,km  \\
Albedo & 0.055  \\
Rotation Period & 9.5\,hours  \\
Semimajor axis & 3.12\,au  \\ 
$F_{\nu}(10\,\mu \rm m)$ & 0.2\,Jy  \\
$F_{\nu}(20\,\mu \rm m)$ & 0.7\,Jy  \\
Surface Temperature & 199$\pm$5\,K\\
Observing date & 2023/4/15  
\enddata
\end{deluxetable}
\vspace{-1.5cm}

\begin{figure*}[ht!]
\centering
\includegraphics[width=18cm]{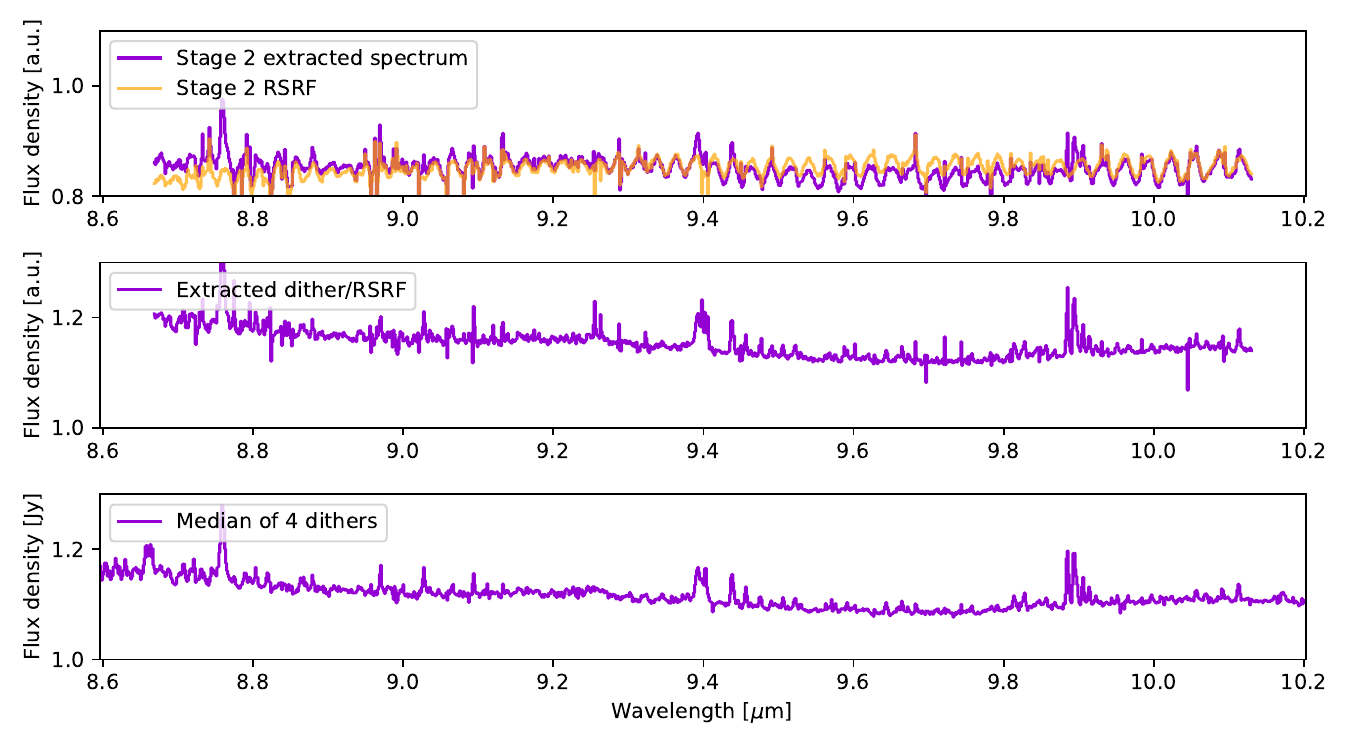}
\includegraphics[width=18cm]{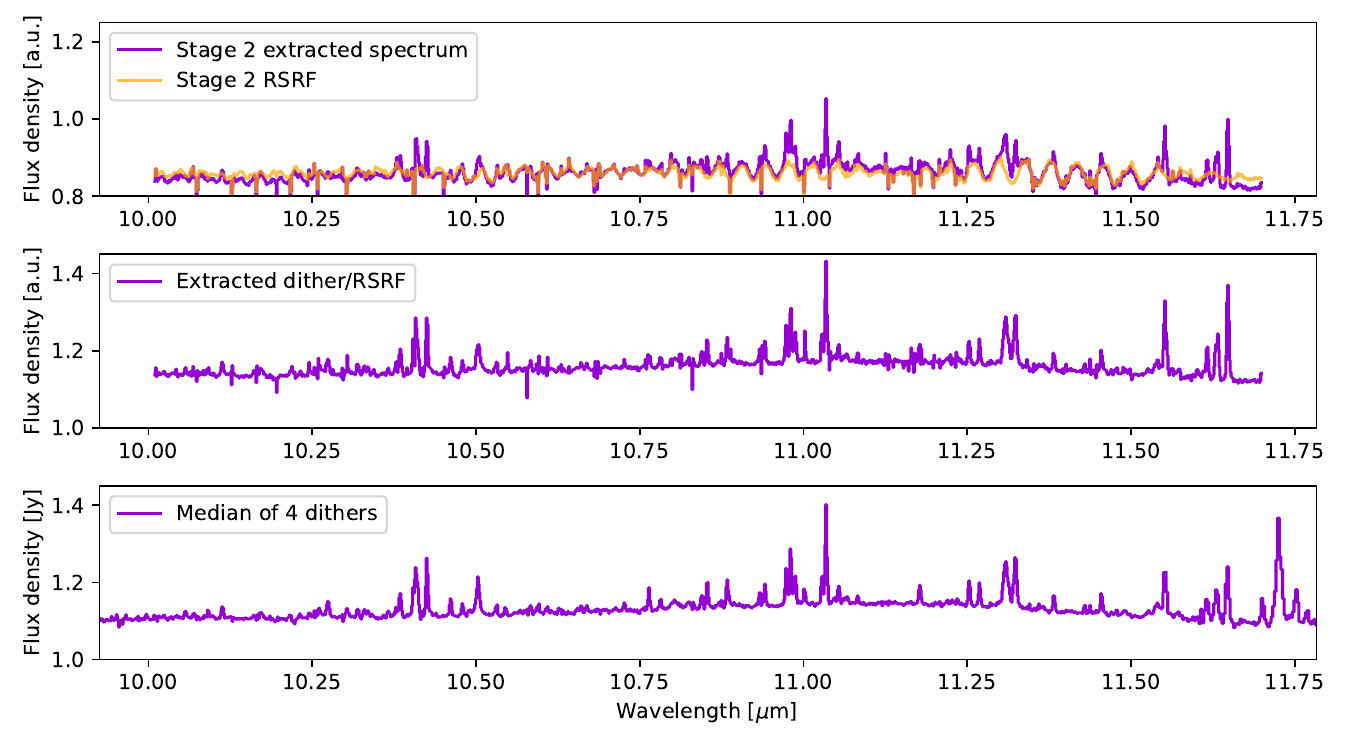}
\caption{Representative example of the empirical RSRF calibration in Channel 2, showing efficient removal of both the MIRI detector fringes, as well as other residual throughput structure. The top panel shows the comparison of the FZ Tau spectrum with that of 526 Jena. The middle panel shows the quality of a single calibrated dither. The bottom panel shows the final median-clipped and stitched spectrum, demonstrating the suppression of outliers. \label{fig:rsrf1}}
\end{figure*}

\begin{figure*}[ht!]
\centering
\includegraphics[width=18cm]{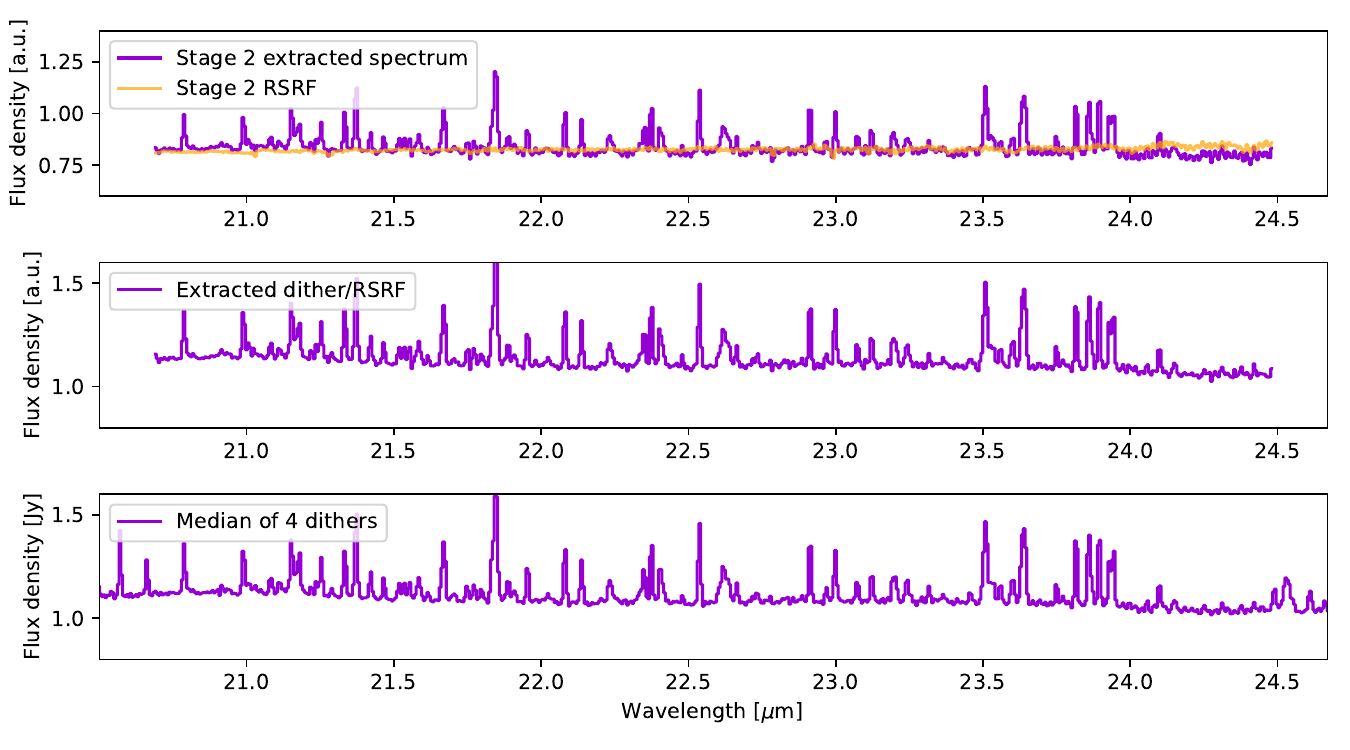}
\includegraphics[width=18cm]{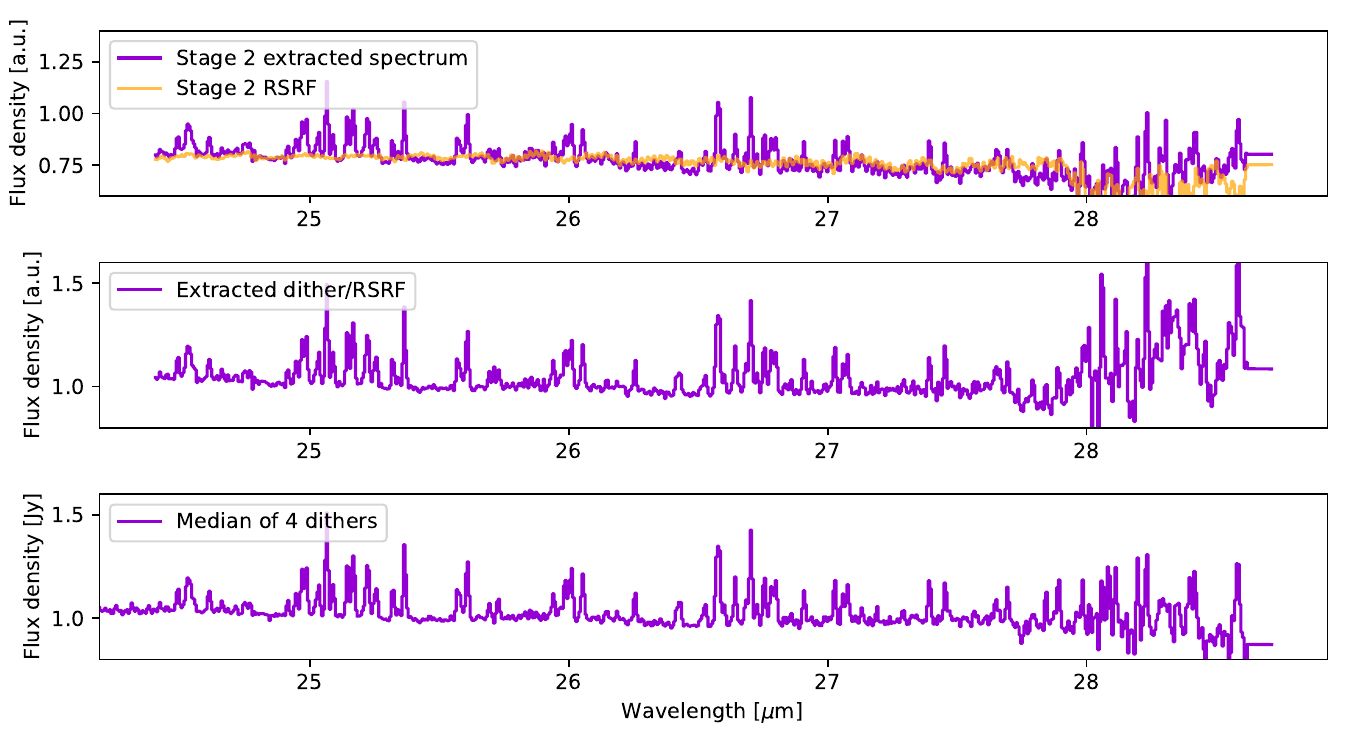}
\caption{Quality of the RSRF in Channel 4. The meanings of lines and colors are the same as those in Figure \ref{fig:rsrf1}. \label{fig:rsrf2}}
\end{figure*}

\section{Data reduction}
\label{section:data_reduction}
\subsection{JDISCS calibration procedure}
The raw FZ Tau and and asteroid JWST-MIRI data from JDISCS used in this paper are available at \dataset[10.17909/7g7e-3w15]{http://dx.doi.org/10.17909/7g7e-3w15}. JDISCS generally uses the latest\footnote{For the FZ Tau spectra shown in this paper, this was pipeline version 1.11.4 and Calibration Reference Data System (CRDS) context 1100.} JWST Calibration Pipeline \citep{Bushouse22} to process the MRS spectra to stage 2b, which produces separate three-dimensional cubes for every exposure, channel, and sub-band. The background is removed by pairwise subtraction of the level 2b cubes obtained at two opposing dither positions. This approach is also efficient in removing many hot pixels from the nod-subtracted cubes. We create a clean, high signal-to-noise two-dimensional image for each channel and sub-band using the median of each cube in the wavelength direction. This enables the use of a Gaussian centroid to determine the exact location of the source. We then extract a one-dimensional spectrum using an aperture that expands linearly with wavelength within each sub-band. Due to defocusing by detector scattering, the MRS PSF is significantly broader than the diffraction limit across the wavelength range, spanning from a factor 2 in Channel 1 short to a factor 1.2 in Channel 4 \citep{Argyriou23}. Consequently, we use extraction diameters of 2.8 times $1.22\times \lambda/D$. We use the same extraction aperture for the target and calibrator, as this negates the need to apply an aperture correction, and the absolute flux calibration precision is essentially that of the JWST pipeline. The exact choice of aperture size does not affect the absolute spectro-photometric calibration under the assumption that both are point sources. 

Each of the four dithered one-dimensional spectra are individually divided by one of four empirical RSRFs from the same position within the MRS Field-Of-View (FOV). The set of one-dimensional dithers are then median-clipped to produce a single one-dimensional spectrum per channel/sub-band, which efficiently removes nearly all residual outliers. Before division, the RSRF is cross-correlated with the spectrum to determine if any small offsets are present; if one is detected, the RSRF is translated in wavelength to match. Typically, we find small wavelength offsets of less than 1/2 resolution element, possibly due to non-reproducibility of the grating position. In Figures \ref{fig:rsrf1} and \ref{fig:rsrf2}, we show the result of this process for a representative set of wavelength ranges. 

\subsection{Refinement of the MRS wavelength calibration}
\begin{figure}[ht!]
\centering
\includegraphics[width=8.5cm]{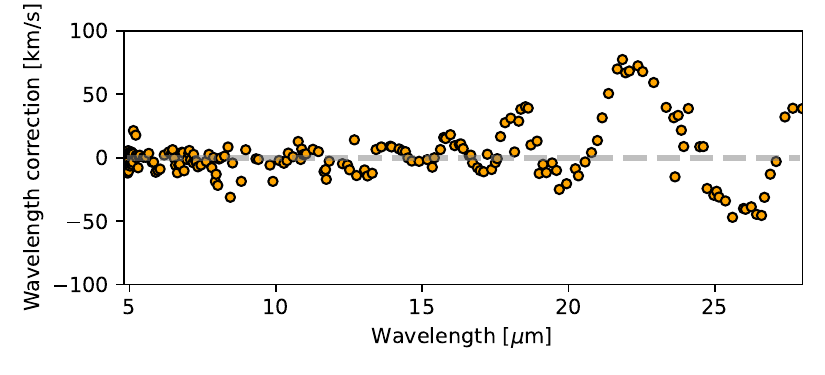}
\caption{Wavelength correction relative to the JWST pipeline at the time of writing (CRDS context 1100) using $\approx 200$ molecular lines. In this context, corrections of up to 90\,$\rm km\,s^{-1}$ are needed to achieve an accuracy of a few\,$\rm km\,s^{-1}$ across the full spectrum. After correction, the intrinsic scatter of the wavelength solution is reduced to $\sim 3-5\,\rm km\,s^{-1}$ (RMS). \label{fig:wavelength correction}}
\end{figure}

Because of the large number of water lines present in mid-IR spectra of planet-forming disks, they provide excellent tests of the MRS wavelength calibration. We use $\approx 200$ single lines or line complexes (clusters of 2--4 nearby lines with separate peaks) from water, CO, and OH to shift the observed FZ~Tau spectrum to the model spectra by cross-correlation. We find that the JWST pipeline wavelength solution in context 1100 has departures of up to 90\,$\rm km\,s^{-1}$ (about 1/2 resolution element at 22\,$\mu$m), with a relatively worse performance for channel 4 (Figure \ref{fig:wavelength correction}). As a result, we derive a new wavelength calibration by assuming the molecular (water and CO) lines trace the systemic velocity of FZ Tau \citep[Heliocentric stellar $R_V = 15.8\pm3.5$\,$\rm km\,s^{-1}$, ][]{Banzatti19}. This assumption is supported by ground-based high-resolution spectroscopy of both the CO fundamental band around 4.7\,$\mu$m and rovibrational and rotational water lines from 5\,$\mu$m to 12.4\,$\mu$m \citep{Banzatti23}.

\subsection{MRS resolving power}
\begin{figure}[ht!]
\centering
\includegraphics[width=9cm]{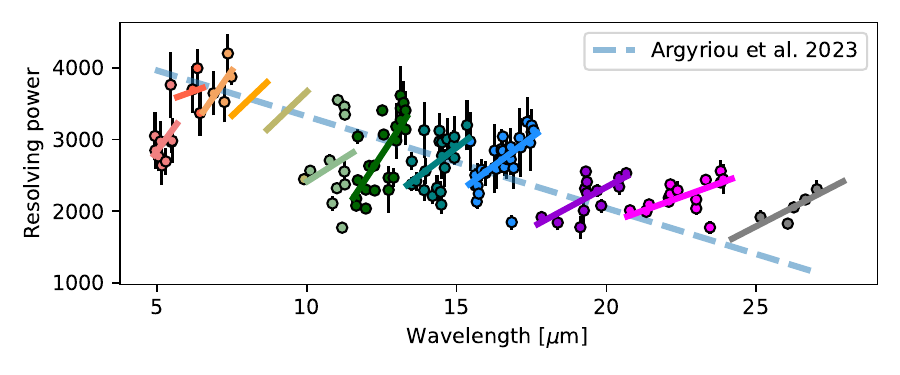}
\caption{Resolving power measured on isolated water and CO lines in the FZ Tau spectrum. The individual measurements have been deconvolved with the intrinsic line width of 36\,$\rm km\,s^{-1}$ (a 7\% difference for $R=3000$). Each MRS sub-band is indicated by a different color, and is fitted separately by a straight line. The relation measured by the MIRI instrument team using a planetary nebula in the Small Magellanic Cloud is also shown \citep{Argyriou23}. For subbands 2A and 2B, we do not find sufficiently suitable lines in the FZ Tau spectrum, so we assume the previous relation and the average subband slope. \label{fig:resolving_power}}
\end{figure}

The large number of unresolved molecular lines provides an opportunity to accurately measure the MRS resolving power, in particular in Channel 4, where previous measurements have lacked suitable lines. We use relatively isolated CO and water lines in the FZ Tau spectrum and fit single Gaussians to measure the FWHM of each line. We deconvolve each MIRI line with the measured intrinsic CO line width of 36\,$\rm km\,s^{-1}$ \citep[from ISHELL, ][]{Banzatti23}. We show the resulting measurements in Figure \ref{fig:resolving_power}. With our measurements we increase the number of resolving power measurements by a factor of several. We detect the slope within 10 out of 12 subbands, as seen in pre-flight data \citep{Glasse15}. Most importantly, we find that the resolving power in Channel 4 is higher than previously thought; The relation of \cite{Argyriou23} decreases to $R\sim 1000$ at the longest wavelengths, when in actuality $R\sim 2-3000$ across all of Channel 4. This has significant consequences when fitting lines in this range, and affects the JDISCS analysis. 

\begin{deluxetable}{lcccc}[ht!]
    \tablecaption{Coefficients of linear fits for each sub-band to the measured MIRI-MRS resolving power ($\lambda/\Delta\lambda = R = A+B\lambda$, with $\lambda$ in $\mu$m), based on the FZ Tau spectrum. The quality column indicates (1) good-quality fit, (2) small number of measurements with large scatter, (3) assumed relation using \cite{Argyriou23} and an average slope.}
\label{table:resolving_power}
\tablehead{
\colhead{Sub-band} & \colhead{A} & \colhead{B} & \colhead{Wavelength} & \colhead{Quality} \\
                   &             &             & $\mu$m               &
}
\startdata
1A & -19.5 & 572 & 4.90--5.74  &  (1)\\ 
1B & 2742  & 150 & 5.66--6.63  &  (2)\\ 
1C & -543  & 601 & 6.53--7.65  &  (2)\\ 
2A & 332   & 400 & 7.51--8.77  & (3)  \\ 
2B & -331  & 400 & 8.67--10.13 & (3) \\ 
2C & -231  & 264 & 10.02--11.70&  (2)\\ 
3A & -5120 & 633 & 11.55--13.47& (1) \\ 
3B & -1871 & 317 & 13.34--15.57& (1) \\ 
3C & -2445 & 312 & 15.41--17.98& (1) \\ 
4A & -2166 & 225 & 17.70--20.95& (1) \\ 
4B & -1176 & 150 & 20.69--24.48& (1) \\ 
4C & -3601 & 216 & 24.19--28.10& (1) 
\enddata
\end{deluxetable}

\begin{figure}[ht!]
\centering
\includegraphics[width=9cm]{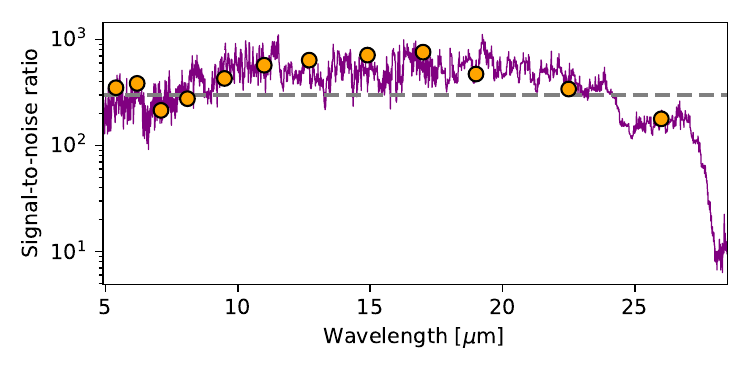}
\caption{Measured signal-to-noise (RMS) achieved with the JDISCS pipeline for the FZ Tau observation (purple curve). The dashed line indicates the expected goal of SNR=300. The orange circles show the prediction by the JWST ETC. Note that the division by the calibrator adds noise, which is included in the ETC prediction. \label{fig:signal-to-noise}}
\end{figure}

\subsection{The 6.1/12.2\,$\mu$m spectral leak}
A serendipitous advantage of asteroid RSRF calibrators is that they are naturally unaffected by the spectral leak of the MRS. It is known from \citet{Gasman23_fringe} that a 2-3\% of the 6.1\,$\mu$m light (channel 1 MEDIUM) leaks into channel 3 SHORT at 12.2\,$\mu$m. Because the asteroid SED is very red, there is a factor 40 between 6.1 and 12.2\,$\mu$m. This effectively decreases the leak to negligible levels in the calibration for spectra reduced using the JDISCS process. 

\subsection{Channel 3 and 4 throughput degradation}
Broadly consistent with reports from the MIRI team, we detect a relative decrease in apparent response compared to the pipeline calibration in channels 3 and 4, with the difference being the largest toward the longest wavelengths. At 27\,$\mu$m (Channel 4C), the signal suggests a further decrease in response from 77\% to 70\% relative to the calibrator between November 21, 2022 and April 15, 2023, which suggests a slowing rate of degradation. At shorter wavelengths, the effect appears to be significantly smaller. Generally, the loss of throughput does not affect the results of this study, or JDISCS in general, besides the associated loss of sensitivity. We do not speculate on the root cause of the throughput loss.  

\subsection{Achievable spectral contrast with the JDISCS pipeline}
\label{section:contrast}
\begin{figure}[ht!]
\centering
\includegraphics[width=9cm]{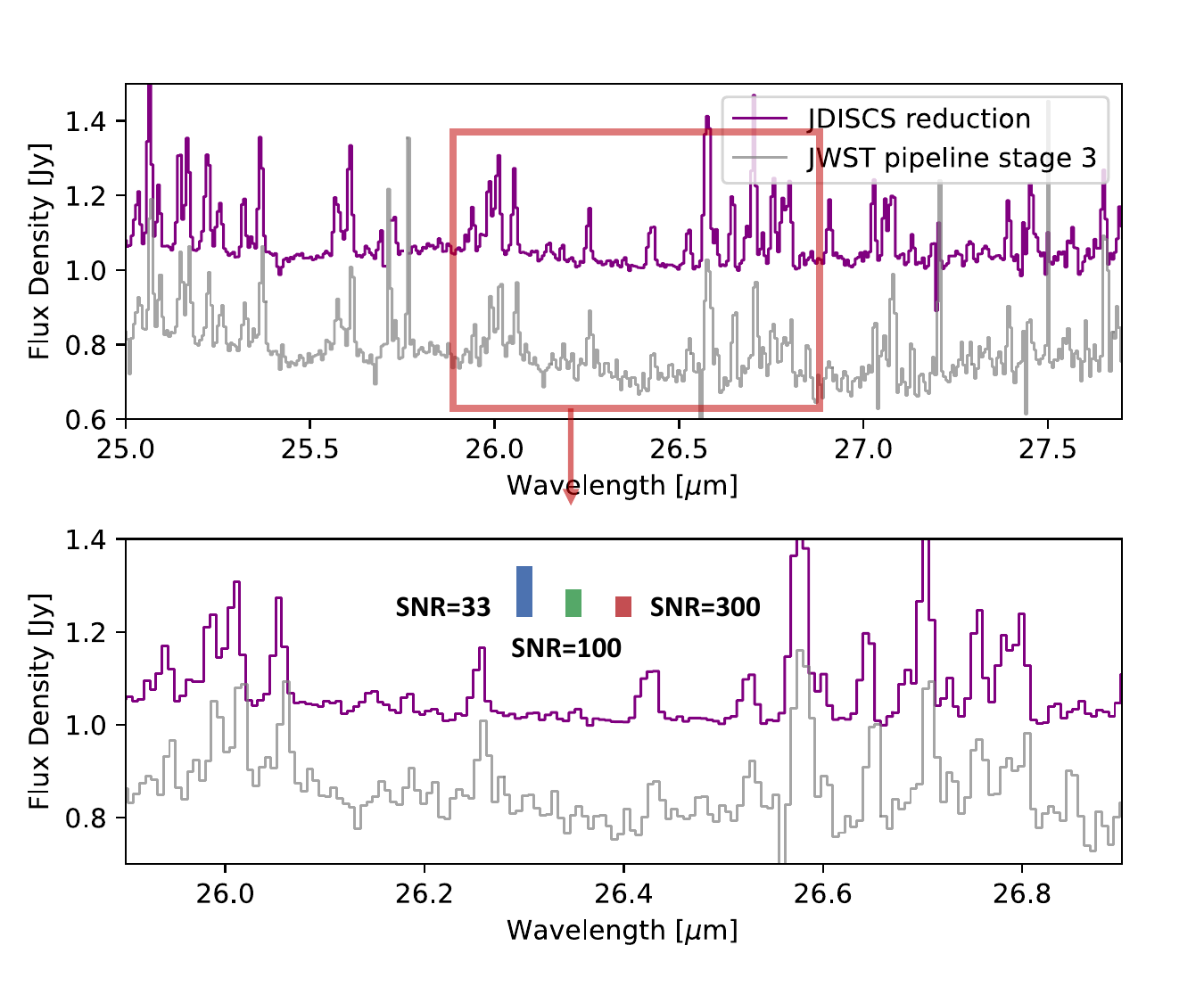}
\caption{Comparison of the JDISCS reduction with the JWST pipeline for a section of Channel 4, which contains the lowest energy water lines. Different levels of contrast (SNR=33, 100, and 300) are indicated by bars showing $\pm 1\sigma$ ranges. In the spectral range shown the JDISCS spectrum reaches $\rm SNR\sim 300$, whereas the pipeline is limited by the calibration to $\rm SNR\sim 50$. The wavelength offsets in the JWST pipeline are apparent around 26\,$\mu$m. \label{fig:pipeline}}
\end{figure}

Figure \ref{fig:signal-to-noise} shows the computed contrast of the FZ Tau spectrum, derived as the root-mean-square (RMS) of the 4 calibrated dithers. This performance is within $\sim$30\% of that predicted by the JWST Exposure Time Calculator, showing that the asteroid-based fringe correction achieves its goal of a contrast better than 1:300 out to 25\,$\mu$m, and is not limited by the fringe correction for sources with a continuum brightness up to at least 1 Jy. The practical difference between JDISCS and the JWST pipeline product for Channel 4C is shown in Figure \ref{fig:pipeline}. 

\section{Analysis}
\label{section:analysis}
\subsection{Dust features}
\label{section:dust}
\begin{figure}[ht!]
\centering
\includegraphics[width=9.cm]{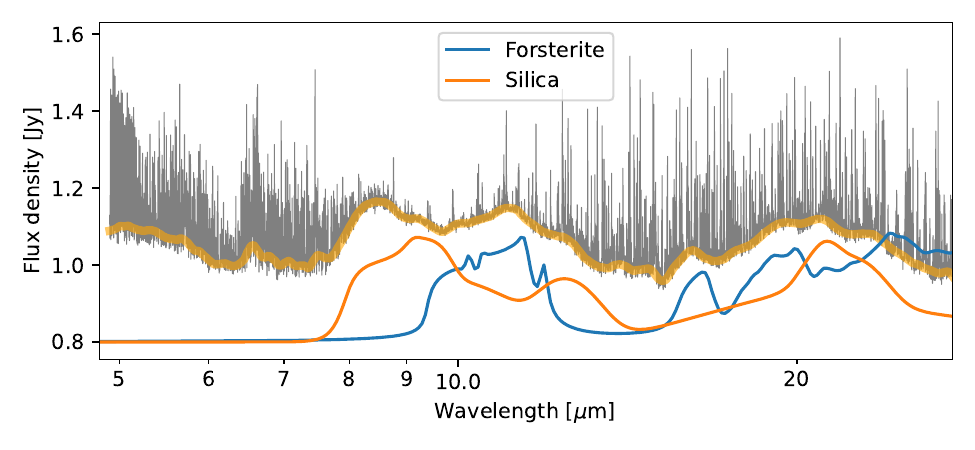}
\caption{Continuum estimate for FZ Tau (orange curve), assuming a pure emission line spectrum, compared to the observed spectrum (grey). Shown are also reference opacities from pure silica and pure forsterite, computed for grain size distributions with $a_{\rm min}=2\,\mu$m, $a_{\rm max}=20\,\mu$m, and $N(a)\propto a^{-3.5}$. The opacities are not fitted to the data, but indicate the locations of significant spectral features. \label{fig:continuum}}
\end{figure}

The FZ Tau continuum is characterized by dust emission from crystalline silicates. To determine the continuum in the presence of many blended emission lines, we use an iterative algorithm\footnote{An implementation of the continuum estimator is available on {\tt https://github.com/pontoppi/ctool}} as follows. The spectrum is median filtered using a wide box of $\sim 100$ wavelength channels. This produces a smoothed spectrum, which can be compared to the previous iteration of the continuum estimator. A new continuum estimate is then constructed by retaining wavelength channels that have lower flux density (assuming emission lines) than the smoothed version of the previous iteration. The rejected wavelength channels are filled in by linear interpolation, producing a new estimate of the continuum, ready for a new iteration. After 3-5 such iterations, a final, nearly noiseless, continuum remains after one final smoothing step using a 2nd-order Savitzky-Golay filter. Note that if there are particularly broad, dense areas of gas-phase emission lines, such as the HCN bands near 14\,$\mu$m, these should be removed from the continuum estimator, as they can otherwise act as a pseudo-continuum. The full FZ Tau spectrum and the continuum are shown in Figure \ref{fig:continuum}.

We compare to representative dust opacity models generated using {\tt optool} \citep{Dominik21}. As seen in Figure \ref{fig:continuum}, the FZ Tau continuum spectrum is dominated by sharp features from a combination of crystalline silicates (in particular forsterite) and silica glass grains with sizes of 1-10\,$\mu$m. This is consistent with the findings by \cite{Sargent09}, based on spectra obtained with the Spitzer InfraRed Spectrometer (IRS). While a detailed fit of the dust continuum is beyond the scope of this paper, the apparent dominance of strong crystalline features indicate significant heating and processing of the inner disk dust in FZ Tau, to a higher degree than those observed in most protoplanetary disks \citep{Kessler-Silacci06, Furlan11}.

\subsection{Molecular lines}
\label{section:lines}
The continuum-subtracted MIRI MRS spectrum of FZ Tau is dominated by bright emission from warm water vapor, as evidenced by both the rovibrational bending mode of water in the 5-8\,$\mu$m range, and pure rotational water lines at longer wavelengths. There is CO present in the fundamental ($\Delta v=1$) band, as seen in the part of the P-branch covered by the MRS range in channel 1A, and numerous lines from warm OH is seen at longer wavelengths in channels 3 and 4. The well-known bands from carbon-bearing molecules C$_2$H$_2$ (13.7\,$\mu$m) and HCN (14.0\,$\mu$m) are weak, while CO$_2$ (15.0\,$\mu$m) is clearly detected (see Figure \ref{fig:spec_long}). While noting the presence of the organics bands, we defer their quantitative analysis to a later JDISCS paper. \citet{Sargent14} reported the detection of gaseous H$_2$CO in absorption near 5.8\,$\mu$m in FZ Tau, but we do not confirm it (see Figure \ref{fig:spec_short}). It is likely that the combination of strong emission lines from the water bending mode and the low resolution of Spitzer-IRS below 10\,$\mu$m mimicked absorption from other species, underlining the critical need for higher spectral resolving power for the study of mid-infrared gas-phase lines.

We extracted lines from the continuum-subtracted spectrum using super-positions of Gaussian profiles to separate line blends. Lines closer than 1$\sigma$ (based on the intrinsic resolving power of MRS) to another line were excluded from the fit to avoid degenerate line strengths. The fits were stabilized by fixing the line centers to their known values from the HITRAN database \citep{hitran22}, and the line widths using the MIRI-MRS resolving power curve measured by JDISCS (Figure \ref{fig:resolving_power}). Consequently only the line amplitudes were varied by the fit, similarly to a procedure applied in the past to de-blend Spitzer-IRS spectra \citep{banzatti13}. 

\begin{figure}[ht!]
\centering
\includegraphics[width=9cm]{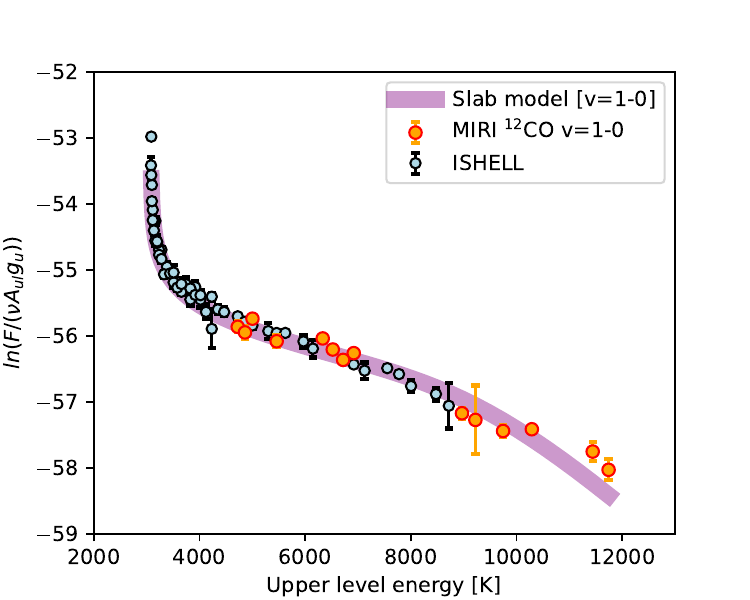}
\includegraphics[width=9cm]{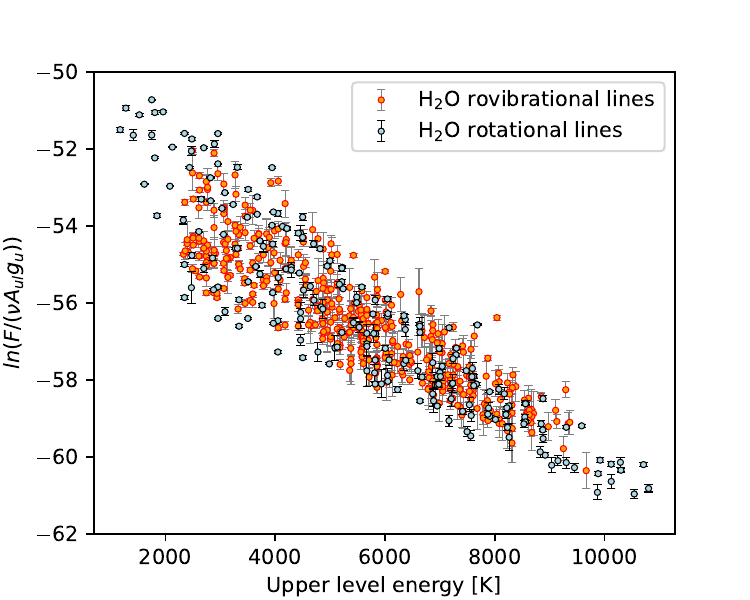}
\caption{Rotation diagrams for the water and CO lines are shown in MKS units, where $F$ is the integrated line flux, $\nu$ is the frequency of the transition in wavenumbers, $A_{ul}$ is the Einstein A coefficient, and $g_u$ is the statistical weight of the upper level. The CO lines detected with MIRI are compared to those measured with IRTF-ISHELL from the ground \citep{Banzatti23}. We find excellent agreement between the ISHELL and MIRI CO lines where there is overlap. The water diagram includes over 600 individual lines \label{fig:rot_diagrams}.}
\end{figure}

\subsection{Physical parameter retrieval}
\label{section:retrieval}
The mid-infrared molecular emission spectrum is an important tracer of processes important for planet formation, such as carbon, oxygen, and nitrogen chemistry, and gas-solid dynamics like pebble drift \citep[e.g.][]{Kalyaan21}. As current instrumentation generally does not spatially resolve the inner disk emission, with a few notable exceptions, we may rely on integrated line fluxes to infer information about the spatial distribution of molecular disk gas. This is possible because of the wide dynamic temperature and opacity range traced by the rich infrared molecular bands. 

\begin{deluxetable}{lccc}[ht!]
\tablecaption{Physical parameters of molecular emission in FZ Tau. Quoted errors are estimated 10\% systematic uncertainties, based on variation on retrieved parameters when varying the selection of lines, different calibrators, etc.
\label{table:slab_parameters}}
\tablehead{
\colhead{Molecule} & \colhead{T} & \colhead{N} & \colhead{A} \\
\colhead{} & \colhead{K} & \colhead{cm$^{-2}$} & \colhead{au$^2$} 
}
\startdata
H$_2$O &516$\pm$50   & (1.6$\pm$0.2) $\times$ 10$^{18}$& 4.52$\pm$1.0  \\
H$_2$O &953$\pm$100  & (2.4$\pm$0.2) $\times$ 10$^{18}$& 0.71$\pm$0.14 \\
CO     &1400$\pm$140 & (3.0$\pm$0.3) $\times$ 10$^{18}$& 0.24$\pm$0.05 
\enddata
\end{deluxetable}

Plane-parallel ``slab'' models provide a convenient way to quantify the properties of the line emission in terms of its excitation temperature, column density, and emitting area. Given the preponderance of evidence that the mid-infrared molecular lines are formed in the inner few au of the disk \citep{Brittain07,Pontoppidan08,Salyk19,Banzatti23}, retrieved physical parameters can be interpreted in the context of a disk surface in Keplerian rotation. We fit the water spectrum at 10--27\,$\mu$m composed of hundreds of rotational emission lines, using a model with one or more independent temperature components, each of which is defined by three parameters that describe the emission from a slab of gas with emitting area $A$, excitation temperature $T$, and column density $N$\citep[e.g.][]{Salyk20}. The model assumes level populations in Local Thermodynamic Equilibrium (LTE), but includes effects from optical depth, including overlapping lines. The intrinsic velocity broading is assumed to be thermal: $\Delta v=\sqrt{k_B T_{\rm gas}/\mu}$, where $T_{\rm gas}$ is the kinetic gas temperature, $\mu$ is the mass of a water molecule, and $k_B$ is the Boltzmann constant. This model is implemented in {\tt spectools-ir} \citep{Salyk22}, and has been used to successfully approximate portions of water spectra observed at lower resolution with Spitzer \citep{Carr11,Salyk11}. {\tt spectools-ir} deploys {\tt emcee} \citep{Foreman-Mackey13} to retrieve posterior distributions of the slab model parameters, given integrated line fluxes, using a Markov-Chain Monte Carlo algorithm. The ro-vibrational bending-mode band of water around 6\,$\mu$m is excluded from the fit because it is known to be affected by non-LTE effects \citep{bosman22,Banzatti23}, but otherwise traces the same range in upper level energies. Indeed, the line strengths of the rovibrational water lines are $\times 4$ weaker than those predicted by fits to the rotational spectrum. This is consistent with the higher critical densities of transitions between vibrational levels. We therefore interpret this as sub-thermal excitation of the vibrational ladder in the bending mode. Since the critical densities for rotational transitions are lower, this is not an indication that the physical parameters retrieved from the rotational lines are similarly out of equilibrium. However, it is important to confirm retrievals using non-LTE water excitation models in future studies.  

\begin{figure*}[ht!]
\centering
\includegraphics[width=14cm]{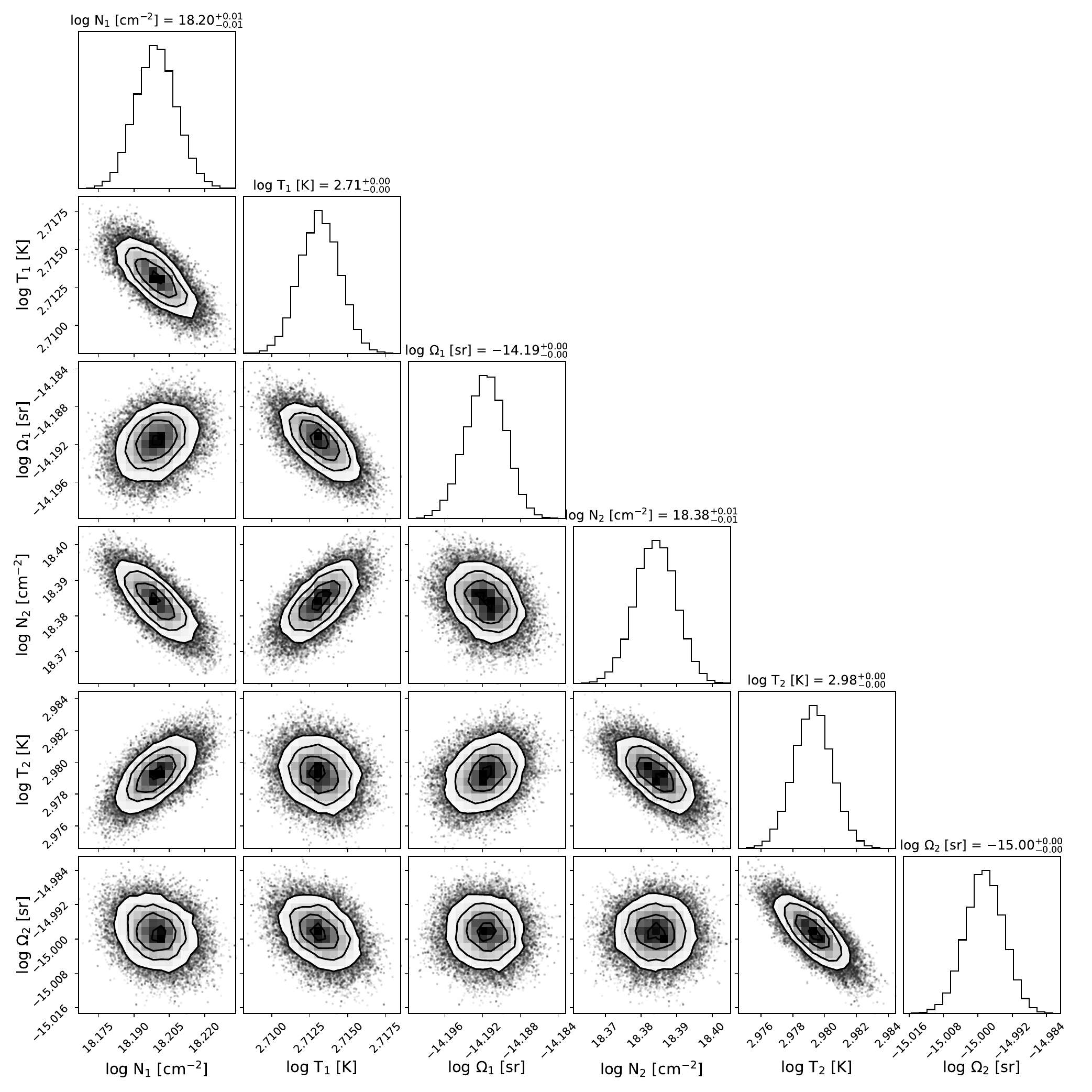}
\caption{Retrieval of a two-component water slab model show that degeneracies are minor, and parameters are statistically well-constrained, supporting the presence of either multiple water temperature components, or a radial temperature gradient. Note that the confidence levels do not include known systematic error, including absolute flux calibration error, errors in the assumed intrinsic line broadening, non-LTE effects, etc. \label{fig:h2o_corner}}
\end{figure*}

Recently, it was found that a single temperature cannot fully reproduce the rotational water lines as observed in disks with MIRI \citep{Banzatti23b}. In the case of FZ Tau, we find that at least two temperature components are required to fit the full range of lines covered by MIRI (about an order of magnitude in upper level energies). However, a single temperature is sufficient to fit the CO v=1-0 band, also when including low-J lines from ground-based observations \citep[][Figure \ref{fig:rot_diagrams}]{Banzatti22}. We show the rotational diagrams of CO and water lines detected in the MIRI spectrum in Figure \ref{fig:rot_diagrams}, and comparisons to slab model calculations. The corner plot for two water temperature components is shown in Figure \ref{fig:h2o_corner}. Slab model parameters for FZ Tau are summarized in Table \ref{table:slab_parameters}.

The water is best fit with temperatures of $T\approx 950$~K and $T\approx 500$~K, similar to the two components previously found in other disks in \citet{Banzatti23b}.  Both components have a column density  of $\gtrsim10^{18}\,\rm cm^{-2}$ in FZ Tau. At these column densities, many water lines are optically thick, providing well-constrained emitting areas. The two components likely reflect an underlying continuous temperature gradient in a surface disk layer. 

\subsection{Extended line emission}
\label{section:extended}
\begin{figure*}[ht!]
\centering
\includegraphics[width=18cm]{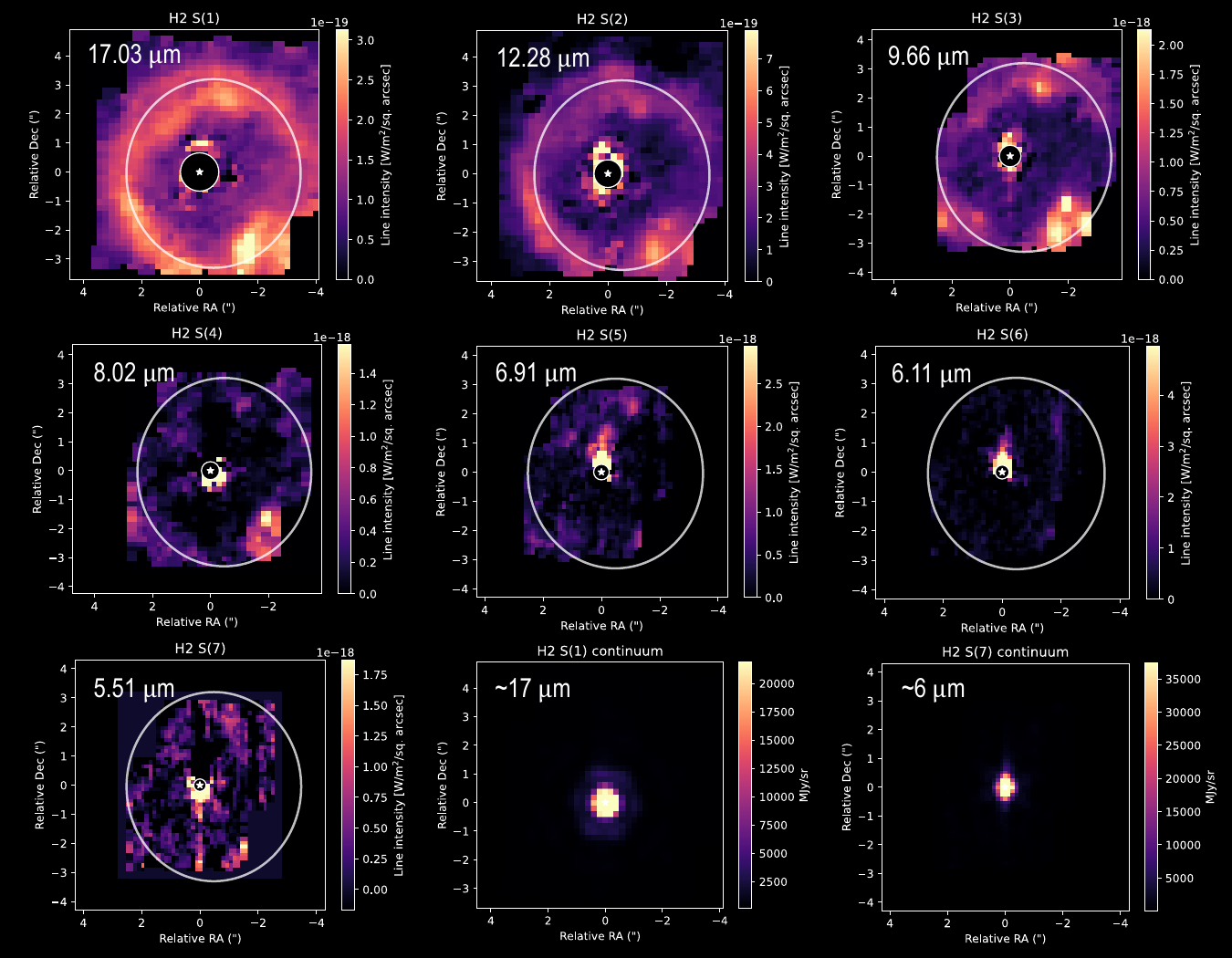}
\caption{Continuum-subtracted line images of the rotational H$_2$ lines. The S(1) and S(2) images show the presence of a ring of cold H$_2$ emission, as well as a number of hot spots to the south-west of the star. The S(3) and S(4) lines are in Channel 2, which has a smaller field of view and therefore, these images do not include the full ring. Further, the Channel 1 field-of-view does not cover the hot spots. The inner working angle with radius of $1.22\lambda/D$ is masked, where $D$ is the JWST telescope diameter, and $\lambda$ is the central wavelength of the line. The location of the star is indicated with a star symbol, and an ellipse is shown outlining the H$_2$ emission ring. \label{fig:h2_ring}}
\end{figure*}

Since MRS provides spatially resolved integral field spectroscopy we conduct a search for extended emission from the detected lines (Figure \ref{fig:h2_ring}). The water lines are spatially unresolved, and therefore confined to an emitting region smaller than the MRS beam (0\farcs175-- 0\farcs45, Half-Width at Half Maximum, corresponding to disk radii of 22.5-58\,au). Since the lines are generally thought to trace inner disk emission from a few au, this is consistent with expectations. 

While undetected toward the central point source of the FZ Tau system, we detect a large extended ring in emission from the rotational H$_2$ ladder. This is particularly apparent in the S(1) and S(2) lines covered by Channel 3 (Figure \ref{fig:h2_ring}). The shape is elliptical with a semi-major axis of $3\farcs25=419$\,au, and a center that is offset from the stellar position by 0\farcs5 along the minor axis. The position angle of the ellipse is consistent with PA=90\degr, but with a large uncertainty. If interpreted as a ring or disk, the axis ratio of the ellipse corresponds to an inclination angle $i\sim 22\degr$, consistent with the submillimeter disk inclination as measured with ALMA of $\sim 25\degr$ (Table \ref{table:fztau_pars}).  

The image further shows the presence of several compact H$_2$ line emission clumps to the south-east of the star. The origin of these is unclear, but they appear to be significantly hotter than the ring itself, with relatively brighter emission in higher-J lines. We do not see them in Channel 1, as the field-of-view is too small. We also detect a feature in H$_2$ S(5)+S(6) and the 6.63\,$\mu$m [NiII] line, resembling a collimated jet oriented roughly north-south and centered on the continuum source. This type of jet is seen in other similar disks \citep{Narang23}, and is consistent with the relatively high accretion rate of FZ Tau. 

While jets are commonly detected with JWST from a range of young stars, the FZ Tau ring is unusual. We cannot uniquely identify its origin, as the line emission is not kinematically resolved with MIRI. However, we consider two different scenarios: The first is that the ring is associated with the FZ Tau disk, and traces material orbiting the star at large radii. The excitation of H$_2$ is potentially complex, and beyond the scope of this paper. Large-scale H$_2$ emission from disks, including debris disks, were seen with the Infrared Space Observatory (ISO) \citep{Thi01a, Thi01b}, and perhaps these are now being recovered by the relatively wide field of MIRI-MRS. A separate scenario is that the ring is part of a wide-angle outflow, viewed close to face-on. This is consistent with the face-on inclination of the disk (Table \ref{table:fztau_pars}), as well as the common occurrence of such outflows from young sources as observed in forbidden emission, including in FZ Tau \citep[e.g.][]{Banzatti19}. However, it is not clear if a face-on outflow can easily produce as a ring-like morphology such as that seen for FZ Tau. 

\section{Discussion}
\label{section:discussion}

The high-contrast, broad-band MIRI spectrum allows us to constrain the spatial distribution of water and other detected species in the disk surface and compare to basic model predictions. We estimate the expected molecular column densities and temperatures from simple considerations. In the inner disk the total dust column density is expected to be highly optically thick, such that only 0.1-1\% of the vertical column is typically contributing to the observed emission lines \citep[e.g.][]{bosman22}. While the inner disk may be depleted in small dust as disks are cleared as part of their natural evolution to the transition disk stage \citep{Zhu12}, this does not appear to have happened yet for FZ Tau \citep{Najita07}. 

\begin{figure}[ht!]
\centering
\includegraphics[width=8.7cm]{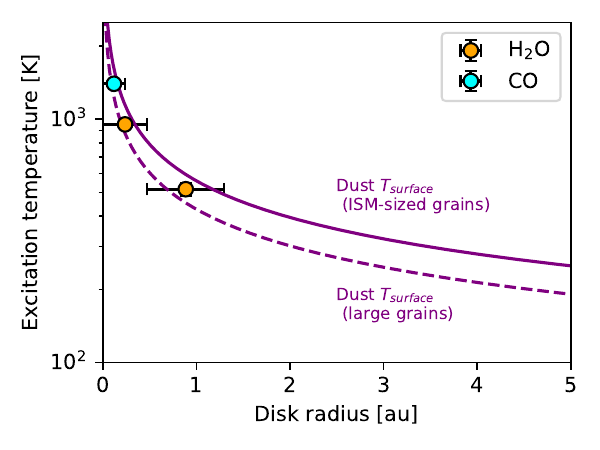}
\includegraphics[width=8.7cm]{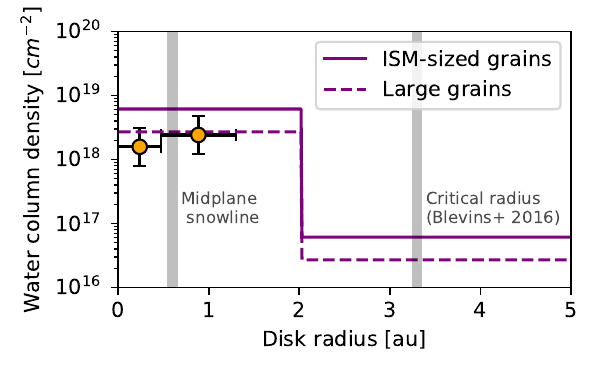}
\caption{Retrieved parameters for the two-component fits of water in FZ Tau, compared to model predictions. Top: The excitation temperatures and disk radii, derived from the slab model emitting areas. The horizontal bars indicate the disk annulus width corresponding to the retrieved emitting areas. The curves indicate surface dust temperatures for FZ Tau using the prescription from \cite{Dullemond01}. Bottom: The retrieved water vapor column densities for the two temperature components. The purple line show the column densities predicted for a water abundance of $n({\rm H_2O})=10^{-4}$ at $T>300\,$K, and  $n({\rm H_2O})=10^{-6}$ at $300\,{\rm K}>T>150\,$K. The critical radius indicates the outer edge of the region with high water abundance region found by \cite{Blevins16} using Herschel upper limits on cool water line fluxes. \label{fig:water_dist}}
\end{figure}

We model the dust temperature in the FZ Tau disk surface using equation 7 from \cite{Dullemond01} and the parameters in Table \ref{table:fztau_pars}. This offers a simple prescription for optically thin reprocessing of starlight in the surface of a flared disk. There is a weak dependence on the ratio of the Planck mean opacities of the stellar irradiation and the disk self-emission, $\sim\epsilon^{-1/4}$. We approximate this from the standard DIANA dust model \citep{Woitke16} to $\epsilon = 0.17$ for disk temperatures of 400-1000\,K and compare to the extreme case of interstellar medium dust \citep[$\epsilon = 0.058$ for $R_V=5.5$;][]{Draine03}. Interpreting the two temperature components as a coarse binning of the temperature gradient, we plot the retrieved gas excitation temperatures in Figure \ref{fig:water_dist} by distributing the emitting areas in consecutive, non-overlapping annuli. The molecular distribution likely extends to larger radii than those observed, but they cannot be well constrained without measuring transitions with upper level energies lower than those covered in the MIRI range ($\lesssim 1000\,$K). We then compare the measurements to the predicted dust surface temperature curve for the FZ Tau parameters (Table \ref{table:fztau_pars}). This provides an excellent match the measured molecular excitation temperate curve, consistent with a scenario in which the molecular gas is strongly coupled to the dust in the line-forming region, and where the excitation of the rotational water lines is well-described by populations in LTE. However, an important caveat is that the effect of superheated gas and subthermal excitation have the opposite effect on line fluxes \citep{Meijerink09}. Thus, we cannot rule out that nature conspires in such a way to make it appear that the lines are coupled to the dust and thermalized. However, gas-dust temperature decoupling can be a substantial effect, increasing the gas temperature factors of 2 or more, depending on disk altitude \citep{Jonkheid04,Glassgold04}. Such departures may be difficult to reconcile with Figure \ref{fig:water_dist}. We also note that previous fits of two-dimensional radiative transfer models to lower-resolution Spitzer data arrived at the opposite conclusion, suggesting gas-dust temperature decoupling and gas-to-dust ratios $\lesssim$100 \citep{Blevins16}. Resolving the question requires further analysis of larger samples of MIRI spectra, such as those offered by JDISCS. 

Similarly, we can consider if the observed column densities are consistent with expectations for the disk surface. The gas is visible above a dust optical depth of $\tau_{\rm dust}=1$ at the wavelength of the relevant transition:

\begin{equation}
N_X = \frac{A_X \times {\rm g2d}}{C_{\rm ext} \mu  m_p},
\label{eq:water_column}
\end{equation}
where $A_X$ is the molecular number abundance relative to hydrogen, $\rm g2d$ is the mass gas-to-dust ratio, $C_{ext}$ is the dust extinction coefficient, $\mu=1.3$ is the mean atomic mass, and $m_p$ is the proton mass. The dust extinction is calculated using {\tt optool} from the DIANA standard dust model parameters\footnote{https://diana.iwf.oeaw.ac.at/} \citep{Woitke16}, truncated to grain sizes $<10\,\mu$m to simulate surface dust, setting $C_{\rm ext}(17\,\mu\rm m)\sim 1800\,\rm cm^2\,g^{-1}$. This choice is driven by evidence than 10\,$\mu$m grains are abundant at the disk surface from SED modeling and mid-infrared scattered light imaging \citep{Pontoppidan07,Duchene23}. Surface grains are unlikely to be much larger. A much smaller grain size would lead to lower opacities. Interstellar medium dust has smaller opacities by about a factor 2, $C_{\rm ext}(17\,\mu\rm m)\sim 791\,\rm cm^2\,g^{-1}$ \citep[$R_V=5.5$,][]{Draine03}, but such small grains are inconsistent theoretical expectations, as well as with observations of disks, as noted above. For a gas-to-dust ratio of 105 \citep{Draine03} and a water abundance of $10^{-4}$ in the inner disk surface \citep{Kamp17}, the water column density predicted by Eq. \ref{eq:water_column} is $N_{\rm H_2O}\sim 2.7\times 10^{18}\,\rm cm^{-2}$. This value is remarkably consistent with the mid-infrared water column densities as estimated in many disks \citep{Salyk11,Banzatti23}. 

Thermo-chemical models without enrichment of water from pebble drift generally predict that the water vapor abundances decrease rapidly below $\lesssim 300\,$K, as gas-phase production of water becomes inefficient \citep[e.g.][]{Kamp17}. In Figure \ref{fig:water_dist} we show the observed column densities along with those predicted for water abundances of $10^{-4}$ at $T>300\,$K and $10^{-6}$ at $T<300\,$K. It is seen that the observed lines in the FZ Tau disk surface are consistent with a high water abundance at least out to 1.5\,AU, corresponding to the 350\,K isotherm. At larger radii, thermo-chemical models predict a significant decrease in column density. However, if efficient drift inwards of icy pebbles is active, this may maintain a high water abundance between 300\,K and 150\,K at larger radii \citep{Kalyaan23,Banzatti23b}. 

The potential enrichment by pebble drift of cold water vapor near ice sublimation at $\sim 150$~K is difficult to test with JWST-MIRI alone, as water lines tracing these temperatures are mostly available at wavelengths $\lambda>30\,\mu$m \citep{Zhang13,Blevins16,Notsu16}. \cite{Blevins16} used observations of low-lying water lines ($E_{\rm upper}=114-800\,$K) with Herschel-PACS to constrain the radial extent of water emission to 3.3\,au, beyond the 300\,K isotherm. For other disks, \cite{Banzatti23b} used the additional wavelength range of Spitzer-IRS to detect cool ($< 300\,$K) water near the snowline, and interpreted its enhancement as evidence for icy pebble drift. 

A second potential signature of pebble drift that JWST-MIRI may address is the absolute abundance of water within the 300\,K isotherm. If this abundance is significantly higher than that predicted by the canonical abundance of oxygen from Eq. \ref{eq:water_column}, this could be evidence for pebble drift enrichment. This effect might be hard to measure when the water emission becomes optically thick and only a surface layer is observed, as the slab results seem to suggest. Finally, the abundance of 300\,K water may correlate with other disk properties linked to pebble drift, such as the size of the dust disk. Evidence for such a connection was presented in \cite{Banzatti23b}, and the compact disk of FZ Tau ($R_{\rm out}=12$\,au) is one where pebble drift is expected to be more efficient. 

Moreover, the water lines in FZ Tau have one of the largest line-to-continuum ratios known \citep{Najita13}, but what underlying property of this disk is causing this observable? Given the predicted water abundance profile in Figure \ref{fig:water_dist}, the inner disk in FZ Tau appears to have filled the region inside the 350\,K isotherm approximating the highest abundance of water possible without additional enrichment of oxygen. If other disks do not maintain a high water abundance ($\sim 10^{-4}$) throughout the inner disk, they might have a smaller line-to-continuum ratio. It is conceivable that this uniformly high water abundance in FZ Tau is regulated by active replenishment by pebble drift. Measurements of the distribution of water in other disks relative to their 300\,K disk surface isotherm may reveal if other disks are less efficient in maintaining a high abundance throughout their inner regions. 

\section{Conclusions}

The calibration methods developed for the JDISC Survey produce high-contrast MIRI spectra across its full wavelength range. We demonstrate SNR$\gtrsim$ 300 and shot-noise limited performance for a test case. We find that the use of bright Themis-family asteroids to measure the RSRF allows us to avoid fringe models to calibrate MIRI spectra. We develop a new MIRI wavelength calibration that is precise to a few km\,s$^{-1}$ across the MIRI range. We generally reproduce the predicted MIRI resolving power, except in channel 4, where we demonstrate higher resolving powers by a factor $\sim$~2--3. This data quality enables the detection of cool water vapor down to 300\,K with MIRI, and more precise retrievals of disk chemistry and gas temperatures. 

We use a MIRI-MRS observation of the FZ Tau protoplanetary disk to test the JDISCS data reduction methods. FZ Tau shows extraordinarily bright molecular emission from water, CO, and OH, it also shows CO$_2$ emission but has weak bands from C$_2$H$_2$ and HCN. Its continuum dust spectrum shows strong bands from crystalline silicates, including forsterite and silica. The molecular emission bands are consistent with a point source at the MIRI MRS resolution, centered on the central star, except for the low-$J$ rotational H$_2$ lines, which show highly extended emission, including a large ring. We speculate that the ring is either part of an outer thin disk for FZ Tau, possibly excited by an external UV radiation field, or is a part of an outflow cone viewed close to face-on. 

The water spectrum of FZ Tau is consistent with an approximately constant water abundance of $\sim 10^{-4}$ in the disk surface at temperatures between $\sim$350 and $\sim$1000\,K, corresponding to radii of 0.2-1.5\,au , for a gas-to-dust ratio of 100 (see Figure \ref{fig:water_dist}). We are not able to constrain the properties of cooler water ($\lesssim 300$\,K) at larger radii, as this would require access to water lines beyond the wavelength coverage of MIRI ($>30\,\mu$m), generally requiring future spectroscopic access to the far-infrared wavelength range \citep{Pontoppidan19}. The excitation temperatures of rotational water lines and CO are consistent with gas coupled to the dust temperature. The high water line contrast observed in FZ Tau does not appear to be linked to exceptionally high column densities, but is more likely an effect of a large emitting area compared to other disks of the same luminosity. We emphasize that this solution is suggestive, but likely not unique. We find that the populations of the vibrationally excited levels of water are subthermal by at least a factor $\sim$4. While the rotational ladder of water appears to be thermalized, we do not rule out that a non-thermal retrieval will yield different gas kinetic temperatures. 

The disk water gradient as measured by MIRI-MRS traces out to a $\sim$~1\,AU, but is ultimately limited by the lack of low-temperature water lines in the MIRI range. To fully map the water distribution in disks across the water snowline requires a new far-infrared space telescope, such as the PRobe Far-Infrared Mission for Astrophysics (PRIMA), \citep{Glenn23,Moullet23}, which proposes to provide high-resolution spectroscopy of the entire thermal range of water vapor.

\facilities{JWST}
\software{{\tt astropy} \citep{astropy22}, {\tt spectools-ir} \citep{Salyk22}}

\begin{acknowledgements}
This research was carried out at the Jet Propulsion Laboratory, California Institute of Technology, under a contract with the National Aeronautics and Space Administration (80NM0018D0004). This work is based on observations made with the NASA/ESA/CSA James Webb Space Telescope. The data were obtained from the Mikulski Archive for Space Telescopes at the Space Telescope Science Institute, which is operated by the Association of Universities for Research in Astronomy, Inc., under NASA contract NAS 5-03127 for JWST. These observations are associated with program 1549. 
\end{acknowledgements}

\appendix

\begin{figure*}[ht!]
\centering
\includegraphics[width=16.7cm]{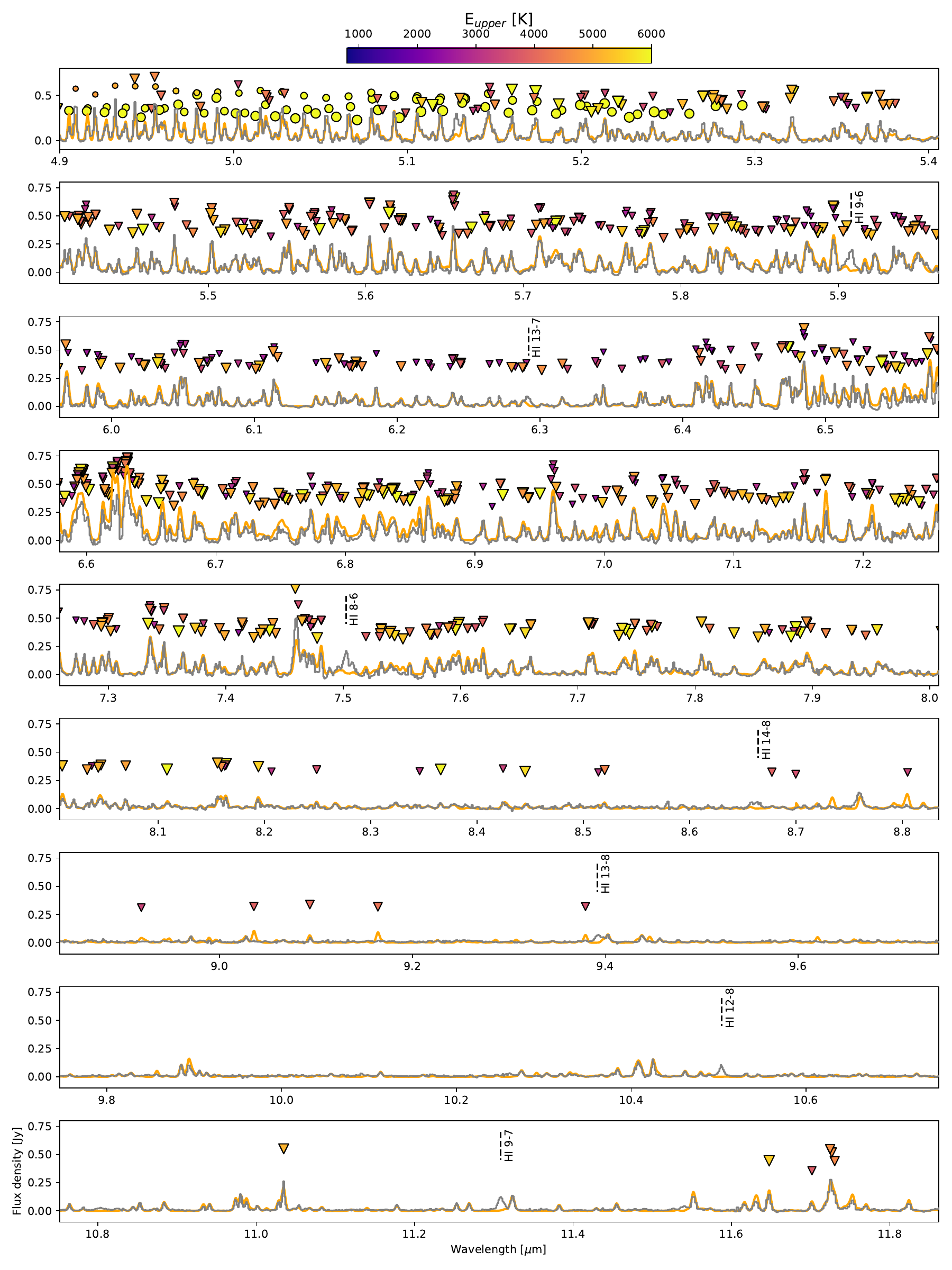}
\caption{Continuum-subtracted emission line spectrum of FZ Tau at $\lambda<11.85\,\mu$m. Bright H$_2^{16}$O lines are marked with triangles,  CO lines with circles, and OH lines with stars. The color and size of the symbols both show the upper level energy of the transition in Kelvin. The orange curve shows the model fit to the spectrum (Table \ref{table:slab_parameters}).  The ro-vibrational lines are not used in the fit, but displayed here at 1/4 of their predicted LTE flux from the model solution in Table \ref{table:slab_parameters}, demonstrating a good match to the rotational populations, although the vibrational levels are sub-thermally populated. Remaining emission is dominated by HI recombination lines. \label{fig:spec_short}}
\end{figure*}

\begin{figure*}[ht!]
\centering
\includegraphics[width=16.7cm]{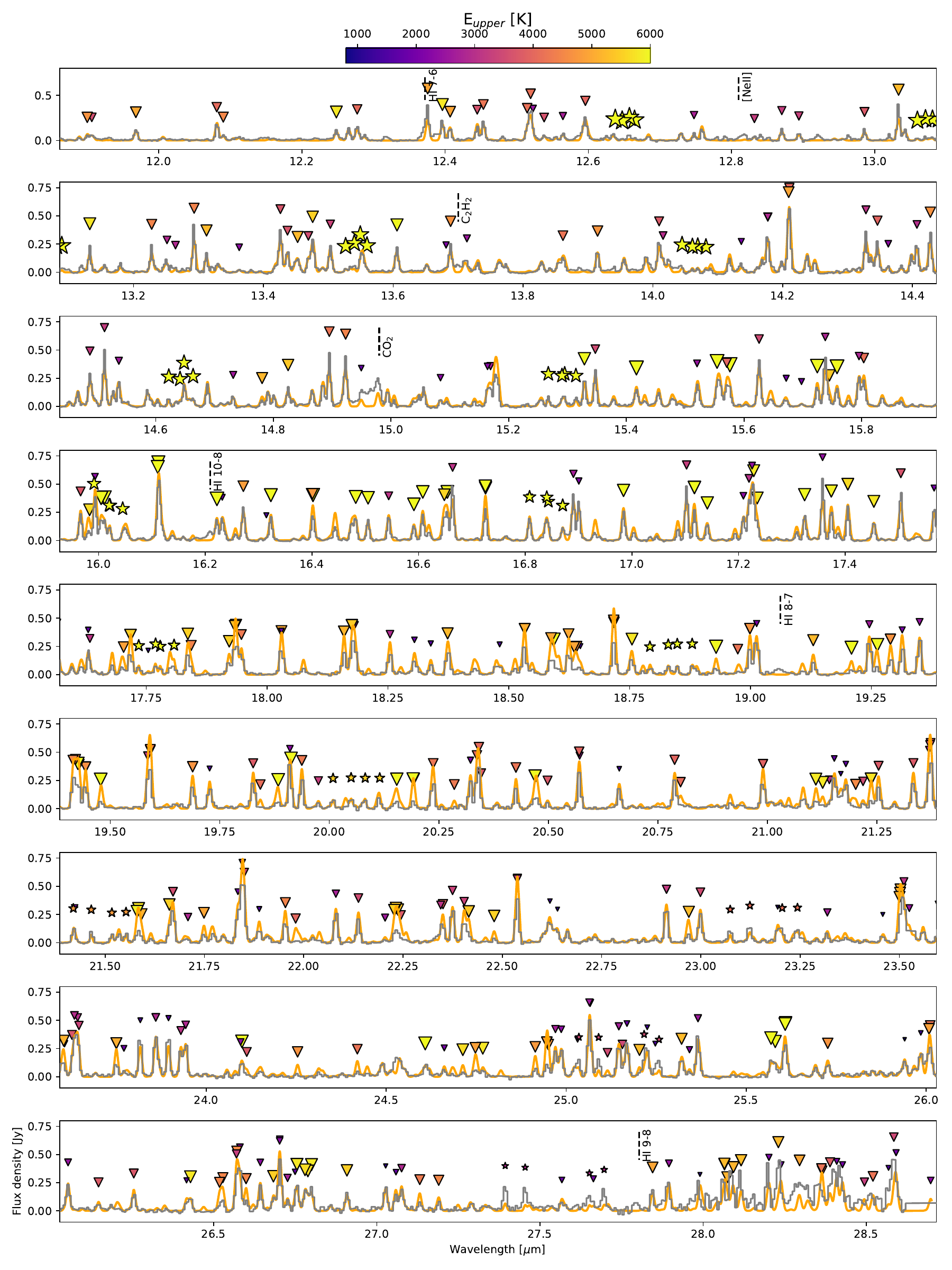}
\caption{Continuum-subtracted emission line spectrum of FZ Tau at $\lambda\ge 11.85\,\mu$m. The labels have the same meaning as those described in Figure \ref{fig:spec_short}. \label{fig:spec_long}}
\end{figure*}

In Figures \ref{fig:spec_short} and \ref{fig:spec_long}, we show the full wavelength range of the continuum-subtracted MIRI spectrum from FZ Tau.

\bibliography{JDISCS}{}
\bibliographystyle{aasjournal}

\end{document}